\documentclass[12pt]{article}

\usepackage{easyReview}
\usepackage{amsmath,amsthm}
\usepackage{amsfonts}
\usepackage{amssymb}
\usepackage{bbm}
\usepackage{mathtools}
\usepackage{mathrsfs}

\usepackage{url}

\usepackage[version=4]{mhchem}
\usepackage{stmaryrd}

\usepackage{subcaption} %
\usepackage{graphicx}
\usepackage{pgfplots}
\usepackage[all]{nowidow}
\usepackage[utf8]{inputenc}
\usepackage{tikz}
\usepackage{multicol}
\usepackage{algpseudocode,algorithm,algorithmicx}
\usepackage{scalerel}
\usepackage{natbib}

\usepackage[normalem]{ulem}

\usepackage[english]{babel}

\usepackage{mathdots}
\usepackage{yhmath}
\usepackage{cancel}
\usepackage{color}
\usepackage{array}
\usepackage{multirow}
\usepackage{gensymb}
\usepackage{tabularx}
\usepackage{booktabs}
\usetikzlibrary{fadings}
\usetikzlibrary{patterns}
\usetikzlibrary{shadows.blur}

\usepackage{enumerate}
\usepackage{algorithm}
\usepackage{algpseudocode}
\usepackage{paralist}

\newtheorem{theorem}{Theorem}

\newtheorem{corollary}{Corollary}

\newtheorem{assumption}{Assumption}

\newtheorem{proposition}{Proposition}
\newtheorem{remark}{Remark}
\newtheorem{definition}{Definition}

\newcommand{\bN}{\mathbb{N}}

\newcommand{\sX}{\mathsf{X}}
\newcommand{\sY}{\mathsf{Y}}
\newcommand{\sZ}{\mathsf{Z}}

\newcommand{\sV}{\mathsf{V}}

\newcommand{\sK}{\mathsf{K}}

\newcommand{\cF}{\mathcal{F}}
\newcommand{\cM}{\mathcal{M}}
\newcommand{\cS}{\mathcal{S}}

\newcommand{\bH}{\mathbb{H}}
\newcommand{\cG}{\mathcal{G}}

\newcommand{\cE}{\mathcal{E}}
\newcommand{\cB}{\mathcal{B}}

\newcommand{\cJ}{\mathcal{J}}
\newcommand{\cN}{\mathcal{N}}

\newcommand{\cP}{\mathcal{P}}
\newcommand{\cD}{\mathcal{D}}
\newcommand{\cH}{\mathcal{H}}

\newcommand{\wt}{\widetilde}
\newcommand{\wh}{\widehat}
\newcommand{\R}{\mathbb{R}}
\newcommand{\E}{\mathbb{E}}
\newcommand{\bP}{\mathbb{P}}
\newcommand{\ones}{\mathbf{1}}
\newcommand{\indic}{\mathbbm{1}}

\DeclareMathOperator{\MMDCI}{MMDCI}

\newcommand{\li}{\langle}
\newcommand{\ri}{\rangle}

\DeclareMathOperator*{\argmin}{arg\,min}

\newcommand{\indep}{\mathrel{\perp\!\!\!\perp}}
\newcommand{\notindep}{\mathrel{\not\!\perp\!\!\!\perp}}

\DeclareMathOperator{\supp}{supp}
\DeclareMathOperator{\diag}{diag}

\usepackage[nodisplayskipstretch]{setspace}

\input{supplement_labels}
\begin{document}
\def\spacingset#1{\renewcommand{\baselinestretch}{#1}\small\normalsize}
\spacingset{1}
\title{\bf Conditional Independence Testing for Random Objects via Generative Models}
\author{Yi Zhang,\hspace{.4cm}Yun Yang\hspace{.2cm}and Xiaofeng Shao
\thanks{Yi Zhang and Xiaofeng Shao, Department of Statistics and Data Science, Washington University in St Louis. Yun Yang, University of Maryland, College Park. Emails: {\tt yiz@wustl.edu}, {\tt yy84@umd.edu} and {\tt shaox@wustl.edu}. Shao's research is partially supported by NSF grant 2524356.}}
\date{\vspace{-3.5em}}
\maketitle
\enlargethispage{-\baselineskip}
\begin{abstract}
We propose a test of conditional independence for random objects taking values in metric spaces of negative type, allowing the conditioning variable to be either a random object or a high-dimensional vector. Our method uses multidimensional scaling and Nystr\"om extensions to obtain finite-dimensional representations of random objects from their pairwise distances, and generative neural networks to estimate conditional distributions in the embedded space. Samples from the fitted generators are then used to construct a doubly robust test statistic. Our main theoretical contribution is to quantify how errors from estimating and truncating the embeddings, together with errors in learning the conditional distributions, affect inference. We establish conditions under which these errors are asymptotically negligible, even when the embedding dimensions grow with the sample size. In particular, we derive explicit convergence rates when the conditional distributions admit generators with smooth, low-dimensional compositional structure. We establish asymptotic validity of the bootstrap test, consistency against fixed alternatives, and nontrivial power against local alternatives. Simulations with distribution-valued data demonstrate its competitive finite-sample performance.
\end{abstract}


\spacingset{1.65}

\section{Introduction}
	
	Modern studies increasingly collect data in the form of complex objects, such as probability distributions, networks, covariance matrices, or functions. Such random objects may take values in metric spaces without a natural linear structure \citep{petersen2019frechet,Dubey2024}, and reducing them to a few scalar summaries can lead to information loss. A central question is whether two such objects remain associated after accounting for other variables. For example, \citet{tan2026dpe} test whether, across countries, the distribution of female age at death is conditionally independent of the distribution of maternal age at childbirth given the distribution of male age at death, which serves as a proxy for the common mortality environment. More generally, one may ask whether an association between two complex measurements persists after adjusting for a large number of covariates. Conditional independence is central to graphical models \citep{pearl2014probabilistic} and variable selection \citep{candes2018panning}, motivating tests that accommodate random objects as the variables of interest, with either object-valued or high-dimensional vector conditioning variables.
	
	Let $X$, $Y$ and $Z$ take values in complete separable metric spaces. We test $H_0:X\indep Y\mid Z$ against general conditional dependence. Unlike Euclidean spaces, these spaces need not have a linear structure or admit reference densities required by many existing tests. When a metric $d$ is of negative type, there exists an embedding into a Hilbert space such that $d(s,t)$ equals the squared distance between the embedded points \citep{lyons2013}. This condition holds for many common object spaces, including Hilbert spaces and univariate distributions under the $2$-Wasserstein distance, and can be extended to more general metric spaces through distance-profile transformations (see Section~\ref{sec:method-population}). Although such an embedding preserves conditional independence, it does not directly yield a practical test. First, the embedding is not explicitly available and must be estimated from pairwise distances between sampled objects. Moreover, since the embedding is typically infinite-dimensional, only finitely many coordinates can be retained. Second, testing conditional independence requires estimating the conditional distributions of the embedded $X$ and $Y$ given the embedded $Z$, a nonparametric problem involving estimated coordinates whose dimensions may grow with the sample size. Since no test can control the type I error over all null distributions while having nontrivial power against all alternatives \citep{ShahPeters2020}, theoretical guarantees require assumptions that make these conditional distributions estimable. This motivates the central question of this paper:
	\begin{quote}
		\emph{Can we construct finite-dimensional representations of random objects from their pairwise distances and learn their conditional distributions accurately enough for valid conditional independence testing? How do errors in estimating these representations and conditional distributions affect inference?}
	\end{quote}
	
	Conditional independence testing has been extensively studied for random vectors. Kernel-based tests characterize conditional dependence through conditional cross-covariance operators in reproducing kernel Hilbert spaces \citep{fukumizu2007kernel,Zhang2011}. Regression-based tests, such as the generalized covariance measure of \citet{ShahPeters2020}, achieve asymptotic validity when the product of the errors of two regression estimators converges to zero sufficiently fast. Generative tests instead use samples from estimated conditional distributions \citep{bellot2019gcit,shi2021dgcit}. In particular, \citet{zhang2024doubly} proposed a maximum mean discrepancy measure of conditional independence (MMDCI) and a doubly robust test based on generative neural networks, which can mitigate the curse of dimensionality by exploiting low-dimensional structure. For Hilbert-space-valued and functional data, \citet{Lundborg2022} extended the generalized covariance measure, while \citet{TangLi2026} developed a kernel test based on a conjoined conditional covariance operator.
	
	For random objects, dependence can be measured using (conditional) distance covariance in metric spaces of negative type \citep{lyons2013,sheng2023distance} or the kernel partial correlation coefficient in general topological spaces \citep{huang2022kpc}. \citet{Dubey2024} developed distance profiles for exploration and inference, while \citet{ZhouMuller2026} introduced profile association and independence tests, with an extension to conditional independence testing for Euclidean conditioning variables based on local linear estimation of conditional distance profiles. More recently, two tests have been developed to accommodate object-valued conditioning variables. \citet{tan2026dpe} embed objects through distance profiles with respect to a reference measure and test conditional independence using a regularized conditional cross-covariance operator. \citet{miyazaki2025sgcm} express the squared norm of this operator in data-dependent spectral coordinates, reducing nuisance estimation to scalar regressions. However, when $Z$ is a random object, both tests rely on kernel ridge regression on $Z$, making their size and power sensitive to the choice of regularization parameter; see Section~\ref{sec:simulation-distributional} for details. Neural networks have also been used for regression with object-valued responses \citep{iao2025deep,zhou2026e2m}, but, to our knowledge, not for estimating their conditional distributions in conditional independence tests.
	
	We propose a generator-based test that combines finite-dimensional representations of random objects with estimation of their conditional distributions. For metric spaces of negative type, the eigenvalues and eigenfunctions of the centered distance kernel define a Hilbert embedding that preserves conditional independence. We estimate its leading coordinates from pairwise distances using classical multidimensional scaling \citep[MDS;][]{torgerson1952multidimensional} and extend the estimated embedding to held-out objects using the Nystr\"om extension \citep{gower1968adding}. We then fit conditional generative neural networks to the estimated coordinates using the kernel loss of engression \citep{shen2025engression}, allowing us to generate samples from the conditional distributions of the embedded $X$ and $Y$ given the embedded $Z$. These samples are used to estimate the conditional mean embeddings in a cross-fitted, doubly robust test statistic based on the MMDCI, which vanishes if and only if $H_0$ holds when characteristic radial kernels are used \citep{ziegel2024characteristic}. We use a wild bootstrap to approximate the null distribution of the statistic. When $Z$ is a high-dimensional Euclidean vector, the generators take $Z$ directly as input without embedding.
	
	Our main contributions are summarized as follows:
	
	\medskip
	\noindent{\bf 1. Conditional independence testing with growing embedding dimensions.} We establish the asymptotic validity of the proposed test, its consistency against fixed alternatives, and nontrivial power against local alternatives of order $n^{-1/2}$, measured by the square root of the population discrepancy (Theorems~\ref{cor:paper-null}--\ref{thm:paper-vector-null}). These guarantees hold even when the embedding dimensions increase with the sample size. Under suitable conditions on the eigenvalue decay, they hold when the conditional distributions admit generators with smooth, low-dimensional compositional structure, allowing the test to mitigate the curse of dimensionality.
	
	\medskip
	\noindent{\bf 2. Error analysis for estimated embeddings and conditional generators.} We derive bounds on the errors in estimating the conditional mean embeddings, accounting for both the MDS embedding and Nystr\"om extension errors and the errors in learning the conditional distributions (Theorem~\ref{thm:paper-generator}). Our analysis reveals two trade-offs. First, increasing the embedding dimensions reduces truncation error but may increase coordinate estimation error due to smaller eigenvalue gaps. Second, a richer generator class can reduce approximation error but may amplify errors in the estimated coordinates of the conditioning variable through a larger Lipschitz constant. Despite these additional errors, the test retains its doubly robust property: under $H_0$, the errors in estimating the two conditional mean embeddings enter multiplicatively, allowing one to converge slowly if the other converges sufficiently fast.
	
	\medskip
	\noindent{\bf 3. Theoretical analysis of MDS embeddings and generative neural networks.} We establish error bounds for the MDS coordinates of training objects and the Nystr\"om coordinates of new objects under a common orthogonal alignment, allowing growing embedding dimensions and requiring only finite second moments of the distances (Theorem~\ref{thm:paper-mds}). We also construct sparse ReLU networks that approximate smooth compositional functions with increasing accuracy while keeping their Lipschitz constants bounded independently of network width (Proposition~\ref{prop:paper-lowdim-approximation}). Together with an analysis of generators trained on estimated coordinates that are identifiable only up to rotation, these results yield explicit convergence rates for the proposed test (Corollary~\ref{cor:paper-lowdim-rates}).
	
	\medskip
	\noindent{\bf Relation to existing work.} Our test builds on the doubly robust MMDCI framework of \citet{zhang2024doubly}, but extending it to random objects requires accounting for errors in estimating and truncating their Hilbert embeddings. Our MDS error bounds extend existing results for bounded kernels \citep{rosasco2010learning} to settings with weaker moment assumptions and growing embedding dimensions, while also covering the Nystr\"om extension to new observations. For neural network approximation, standard deep ReLU constructions typically control the number of parameters rather than the Lipschitz constant, which may grow as the approximation error decreases \citep{jiao2023approximation}. Width-independent norm bounds are known for shallow networks approximating Barron-type functions \citep{klusowski2018approximation,e2022barron}, with related results for their compositions \citep{e2020banach}. Our construction extends this type of control to sparse deep networks approximating smooth compositional functions, allowing the approximation error to decrease without increasing the Lipschitz constant as the network width grows. This property is important for controlling how errors in the estimated embeddings propagate through the generators. Our simulations with distribution-valued data show that the proposed test maintains empirical size close to the nominal level as the dimension of the conditioning variable increases. In contrast, the tests of \citet{tan2026dpe} and \citet{miyazaki2025sgcm} exhibit increasing size distortions and are sensitive to the choice of regularization parameter.
	
	The rest of the paper is organized as follows. Section~\ref{sec:methodology} introduces the Hilbert embedding, its estimation by MDS and the Nystr\"om extension, the conditional generators, and the proposed test. Section~\ref{sec:theory} presents the theoretical results, Section~\ref{sec:simulation} reports the simulations, and Section~\ref{sec:conclusion} concludes. Proofs and implementation details are provided in the supplement.
	
	\medskip
	\noindent{\bf Notation.} Let $\bN_0=\{0,1,2,\ldots\}$ and, for a positive integer $n$, write $[n]=\{1,\ldots,n\}$. For vectors and matrices, $\|\cdot\|_0$ denotes the number of nonzero entries. For a vector $a$, let $\|a\|_\infty=\max_j|a_j|$ and write $[a]_j=a_j$ for its $j$th entry. For a vector-valued function $f$, let $[f]_j$ denote its $j$th component, so that $[f]_j(x)=[f(x)]_j$. We use $\li\,\cdot\,,\,\cdot\,\ri_{\bH}$ and $\|\cdot\|_{\bH}$ to denote the inner product and induced norm of a Hilbert space $\bH$, respectively. We omit the subscript when $\bH$ is a Euclidean space $\R^d$ or the space of square-summable sequences $\ell^2=\{a\in\R^\infty:\sum_{j=1}^\infty a_j^2<\infty\}$, or when the underlying space is clear from the context. For Hilbert spaces $A$ and $B$, let $A\otimes B$ denote their tensor product. For a complete separable metric space $\cS$, let $\cM(\cS)$, $\cM_0(\cS)=\{u\in\cM(\cS):u(\cS)=0\}$, and $\cM_+(\cS)$ denote the spaces of finite signed Borel measures, finite signed Borel measures with zero total mass, and finite nonnegative Borel measures on $\cS$, respectively.

	\section[Hilbert Embeddings and Conditional Independence Testing]{Hilbert Embeddings and Conditional\\ Independence Testing}\label{sec:methodology}
	
	Let $X$, $Y$, and $Z$ be random elements defined on a common probability space
	$(\Sigma,\cF,\lambda)$, taking values in complete separable metric spaces
	$(\sX,d_X)$, $(\sY,d_Y)$, and $(\sZ,d_Z)$, respectively.
	For a metric space $\sV$, let $\cB(\sV)$ denote its Borel sigma-algebra.
	We test $H_0:X\indep Y\mid Z$ against $H_1:X\notindep Y\mid Z$.
	Under $H_0$, for every $A\in\cB(\sX)$ and $B\in\cB(\sY)$,
	\[
	\bP(X\in A,Y\in B\mid Z)
	=\bP(X\in A\mid Z)\bP(Y\in B\mid Z)
	\quad\text{almost surely}.
	\]
	
	Our testing procedure extends the doubly robust MMDCI test of
	\citet{zhang2024doubly} from random vectors to random objects.
	We first embed the random objects into Hilbert spaces in a way that
	preserves conditional independence and retain finitely many coordinates
	(Section~\ref{sec:method-population}).
	We estimate these coordinates from pairwise distances using classical
	multidimensional scaling (MDS; Section~\ref{sec:method-mds}).
	We then fit generative neural networks (generators) to estimate the
	conditional distributions of the embedded $X$ and $Y$ given the
	embedded $Z$ (Section~\ref{sec:method-generators}).
	The fitted generators are used to construct a doubly robust test
	statistic, whose null distribution is approximated by a wild bootstrap
	(Section~\ref{sec:method-statistic}).
	
\begin{figure}[t]
		\centering
		\includegraphics[width=0.85\textwidth]{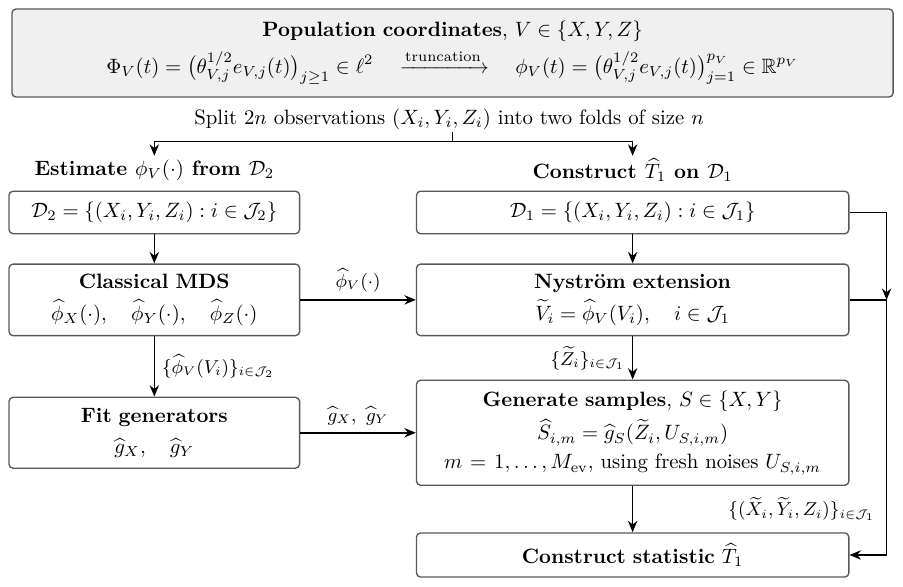}
	\caption{Construction of the statistic $\wh T_1$.}
		\label{fig:method-procedure}
	\end{figure}

	To separate estimation from testing, we use sample splitting and
	cross-fitting. Suppose we observe $2n$ independent copies
	$\{(X_i,Y_i,Z_i)\}_{i=1}^{2n}$, where $n\ge2$.
	Partition $[2n]$ into two disjoint sets $\cJ_1$ and $\cJ_2$,
	each of size $n$, and let
	$\cD_k=\{(X_i,Y_i,Z_i):i\in\cJ_k\}$ for $k=1,2$.
	Specifically, we estimate the MDS embeddings and generators on
	$\cD_2$ and use them to construct a statistic $\wh T_1$ on $\cD_1$.
	The Nystr\"om extension maps the held-out observations into the same
	coordinate system as the training observations, accounting for the
	rotational ambiguity of MDS.
	Exchanging the roles of $\cD_1$ and $\cD_2$ gives a second statistic
	$\wh T_2$, and the two are combined to form the final test statistic.
	Figure~\ref{fig:method-procedure} illustrates the construction of
	$\wh T_1$.
	
	\subsection{Population Hilbert embedding and truncation}\label{sec:method-population}
	
	For each $V\in\{X,Y,Z\}$, assume that $(\sV,d_V)$ is of negative type
	and $\E[d_V(t_0,V)]<\infty$ for some $t_0\in\sV$.
	Recall that a metric space $(\sV,d_V)$ is of negative type if,
	for any $m\ge2$, $t_1,\ldots,t_m\in\sV$, and
	$a_1,\ldots,a_m\in\R$ satisfying $\sum_{i=1}^m a_i=0$, we have
	$\sum_{i,j=1}^m a_i a_j d_V(t_i,t_j)\le0$
	\citep[Equation~(3.1)]{lyons2013}.
	By Schoenberg's embedding theorem \citep{lyons2013}, negative type is
	equivalent to the existence of a separable Hilbert space $\bH_V$ and
	a map $\Phi_V:\sV\to\bH_V$ such that
	$d_V(s,t)=\|\Phi_V(s)-\Phi_V(t)\|_{\bH_V}^2$ for $s,t\in\sV$.
	Since $d_V$ is a metric, $\Phi_V$ is continuous and injective.
	By Corollary~15.2 of \citet{kechris1995classical}, $\Phi_V$
	is a Borel isomorphism onto its image, so
	$\sigma\{\Phi_V(V)\}=\sigma(V)$. Consequently,
	\begin{equation}\label{eq:method-embedded-null}
		X\indep Y\mid Z
		\quad\Longleftrightarrow\quad
		\Phi_X(X)\indep\Phi_Y(Y)\mid\Phi_Z(Z).
	\end{equation}
	Thus, the Hilbert embedding preserves conditional independence,
	allowing us to test the equivalent hypothesis in Hilbert spaces.
	
	The negative type assumption covers many common object spaces,
	including Hilbert spaces, univariate probability distributions under
	the $2$-Wasserstein distance, and positive definite covariance matrices
	under the log-Euclidean distance \citep{zhangzhushao2025}.
	Moreover, if $d_V$ is of negative type, so is $d_V^s$ for any
	$0<s<1$ \citep{lyons2021second}.
	The framework can also be extended to more general metric spaces.
	For an arbitrary metric space $(\sV,d_V)$, let $\varrho_V$ be a
	probability measure with full support on $\sV$. Provided the integral
	is finite, the distance-profile metric
	$d_{\varrho_V}(t,t')=
	\left[\int_{\sV}\{d_V(u,t)-d_V(u,t')\}^2\,d\varrho_V(u)\right]^{1/2}$
	is of negative type \citep{tan2026dpe}.
	Thus, even when the original metric is not of negative type,
	a suitable transformation of the distances can provide a Hilbert
	embedding that preserves the information in the random objects.
	
	Although the embedding preserves conditional independence, testing
	the equivalent null hypothesis in Equation~\eqref{eq:method-embedded-null}
	remains challenging. Existing tests generally require estimating
	functionals of the conditional distributions of $\Phi_X(X)$ and
	$\Phi_Y(Y)$ given $\Phi_Z(Z)$. Without parametric assumptions,
	these estimates may converge slowly and lead to substantial size
	distortion \citep{zhang2024doubly}, particularly when the conditioning
	variable is high- or infinite-dimensional. To mitigate this difficulty,
	we use generative neural networks that can exploit low-dimensional
	structure in conditional distributions under suitable smoothness
	conditions \citep{chen2020distribution}. Samples from the fitted
	generators can then be used to estimate the required conditional
	expectations by Monte Carlo \citep{zhou2022deep}.
	Since these networks require finite-dimensional inputs and outputs,
	we seek an embedding whose coordinates can be estimated from pairwise
	distances and truncated to a finite dimension.
	
	We construct such an embedding using the population version of
	classical MDS \citep[Section~3.2]{lim2024classical}.
	Let $\bP_V$ denote the distribution of $V$, and let $V'$ be an
	independent copy. Define the centered distance kernel $\kappa_V$
	and its integral operator $\sK_V$ on $L^2(\bP_V)$ by
	\begin{equation}\label{eq:method-centered-kernel}
		\begin{aligned}
			\kappa_V(s,t)&=-\frac12\bigl\{
			d_V(s,t)-\E[d_V(s,V)]-\E[d_V(t,V)]
			+\E[d_V(V,V')]\bigr\},\\
			(\sK_V f)(s)&=\int_{\sV}\kappa_V(s,t)f(t)\,d\bP_V(t).
		\end{aligned}
	\end{equation}
	By Lemma~\ref{lem:centered-distance-kernel} in
	Appendix~\ref{app:auxiliary-lemmas}, $\kappa_V$ is continuous and
	positive semidefinite under the negative type assumption.
	Moreover, the triangle inequality implies
	$\E[\kappa_V(V,V)]=\tfrac12\E[d_V(V,V')]
	\le \E[d_V(t_0,V)]<\infty$.
	Hence $\sK_V$ is a positive trace-class operator with eigenpairs
	$(\theta_{V,j},e_{V,j})$, where
	$\theta_{V,1}\ge\theta_{V,2}\ge\cdots\ge0$
	\citep[Theorem~6.1(iii) and (iv)]{devito2013mercer}.
	By Mercer's theorem \citep[Theorem~4.1]{devito2013mercer}
	(see also Lemma~\ref{lem:appendix-mercer} in
	Appendix~\ref{app:auxiliary-lemmas}),
	\[
	\kappa_V(s,t)=\sum_{j\ge1}
	\theta_{V,j}e_{V,j}(s)e_{V,j}(t)
	\quad\text{for }s,t\in\supp(\bP_V),
	\qquad \sum_{j\ge1}\theta_{V,j}<\infty.
	\]
	
	Choose a truncation level $p_V\in[n]$ satisfying
	$\theta_{V,p_V}>\theta_{V,p_V+1}$.
	We define the full population embedding $\Phi_V$ and its
	$p_V$-dimensional truncation $\phi_V$ by
	\begin{equation}\label{eq:method-population-coordinates}
		\Phi_V(t)=\bigl(\theta_{V,j}^{1/2}e_{V,j}(t)\bigr)_{j\ge1}
		\in\ell^2,
		\qquad
		\phi_V(t)=\bigl(\theta_{V,j}^{1/2}e_{V,j}(t)\bigr)_{j=1}^{p_V}
		\in\R^{p_V}.
	\end{equation}
	The full embedding satisfies
	$d_V(s,t)=\|\Phi_V(s)-\Phi_V(t)\|_{\ell^2}^2$
	for $s,t\in\supp(\bP_V)$, and hence preserves conditional
	independence as in Equation~\eqref{eq:method-embedded-null}.
	By Lemma~\ref{lem:paper-mds-coordinate-identities}
	(\ref{item:paper-mds-population}) in Appendix~\ref{app:proof-paper-mds},
	$\E[\Phi_V(V)]=0$, and its coordinates are uncorrelated with
	variances $\theta_{V,1}\ge\theta_{V,2}\ge\cdots$.
	Thus, $\phi_V(V)$ consists of the first $p_V$ principal component
	scores of $\Phi_V(V)$, while the discarded coordinates have total
	variance $\sum_{j>p_V}\theta_{V,j}$, which contributes to the
	truncation error in Section~\ref{sec:theory}.
	
	For a positive integer $r$, define the linear operator
	$J_r:\R^r\to\ell^2$ and its adjoint $J_r^*:\ell^2\to\R^r$ by
	$J_r(a)=(a_1,\ldots,a_r,0,0,\ldots)^{\top}$ and
	$J_r^*(s)=(s_1,\ldots,s_r)^{\top}$
	for $a\in\R^r$ and $s\in\ell^2$. Then
	$\phi_V=J_{p_V}^*\Phi_V$.
	The eigengap condition
	$\theta_{V,p_V}>\theta_{V,p_V+1}$ ensures that the subspace
	spanned by the first $p_V$ eigenfunctions is uniquely determined.
	However, the eigenfunctions are identifiable only up to signs
	and, for repeated eigenvalues, orthogonal transformations within
	the corresponding eigenspaces. Consequently, $\phi_V$ is defined
	only up to an orthogonal transformation of $\R^{p_V}$.

	\subsection{Empirical MDS and Nystr\"om extension}\label{sec:method-mds}
	
	The population coordinates $\phi_V(V_i)$ depend on the unknown
	distribution $\bP_V$. We estimate them from pairwise distances
	using classical MDS \citep{torgerson1952multidimensional} on
	the training sample $\cD_2$, and extend the estimated embedding
	to held-out observations in $\cD_1$ using the Nystr\"om extension
	\citep{gower1968adding}.
	
	For each $V\in\{X,Y,Z\}$, replacing $\bP_V$ in
	Equation~\eqref{eq:method-centered-kernel} with the empirical
	distribution of $\{V_i:i\in\cJ_2\}$ gives, for $s,t\in\sV$, the empirical centered
	distance kernel
	\begin{equation}\label{eq:method-sample-kernel}
		\wh\kappa_V(s,t)=-\frac12\Bigl\{
		d_V(s,t)-\frac1n\sum_{i\in\cJ_2}d_V(s,V_i)
		-\frac1n\sum_{i\in\cJ_2}d_V(t,V_i)
		+\frac1{n^2}\sum_{i,j\in\cJ_2}d_V(V_i,V_j)\Bigr\},
	\end{equation}
	which is positive semidefinite by
	Lemma~\ref{lem:centered-distance-kernel}.
	List the indices in $\cJ_2$ as $i_1,\ldots,i_n$, and let
	$\ones_n$ denote the vector of $n$ ones. Define
	\begin{equation}\label{eq:method-mds-matrices}
		D_V=\bigl\{d_V(V_{i_r},V_{i_s})\bigr\}_{r,s=1}^n,
		\quad H_n=I_n-\frac1n\ones_n\ones_n^{\top},
		\quad B_V=-\frac12H_nD_VH_n.
	\end{equation}
	Then $B_V=\{\wh\kappa_V(V_{i_r},V_{i_s})\}_{r,s=1}^n$
	is positive semidefinite. Let
	$\wh\ell_{V,1}\ge\cdots\ge\wh\ell_{V,p_V}\ge0$ be its leading
	$p_V$ eigenvalues, and let $\Xi_V\in\R^{n\times p_V}$
	contain the corresponding orthonormal eigenvectors as columns.
	Define
	$\wh\Lambda_V=\diag(\wh\ell_{V,1},\ldots,\wh\ell_{V,p_V})$ and
	$A_V=\Xi_V\wh\Lambda_V^{1/2}$.
	The MDS coordinates of $V_{i_r}$ are given by the $r$th row
	of $A_V$, denoted by $A_V(r,\cdot)^{\top}$.
	Since $B_V/n$ is the empirical counterpart of the integral
	operator $\sK_V$, its eigenvalues $\wh\ell_{V,j}/n$ and
	eigenvectors $\sqrt n(\Xi_V)_{\cdot j}$ provide empirical
	analogues of $\theta_{V,j}$ and $e_{V,j}$, respectively.
	Hence $A_V(r,\cdot)^{\top}$ estimates the population
	coordinates $\phi_V(V_{i_r})$, up to an orthogonal transformation.
	
	To extend the estimated embedding beyond the training sample,
	let $\wh\Lambda_V^{\dagger/2}$ be the diagonal matrix whose
	$j$th diagonal entry is $\wh\ell_{V,j}^{-1/2}$ if
	$\wh\ell_{V,j}>0$ and zero otherwise.
	The Nystr\"om extension defines the coordinates of any $t\in\sV$ by
	\begin{equation}\label{eq:method-nystrom}
		\wh\phi_V(t)
		=\wh\Lambda_V^{\dagger/2}\Xi_V^{\top}
		\bigl(\wh\kappa_V(V_{i_r},t)\bigr)_{r=1}^n
		\in\R^{p_V}.
	\end{equation}
	This extension agrees with the MDS coordinates on the
	training sample, since
	$\wh\phi_V(V_{i_r})=A_V(r,\cdot)^{\top}$ for $r\in[n]$.
	To see its connection with the population embedding, note that
	$\sK_Ve_{V,j}=\theta_{V,j}e_{V,j}$ implies
	$[\phi_V(t)]_j=\theta_{V,j}^{-1/2}
	\int_{\sV}\kappa_V(t,s)e_{V,j}(s)\,d\bP_V(s)$
	for $j\in[p_V]$ and $t\in\supp(\bP_V)$.
	Replacing the population quantities in this identity with their
	empirical counterparts yields Equation~\eqref{eq:method-nystrom}.
	
	For $i\in\cJ_1$, write $\wt V_i=\wh\phi_V(V_i)$ for the estimated
	coordinates obtained from $\cD_2$.
	The empirical embedding $\wh\phi_V$ estimates $\phi_V$ up to a
	data-dependent orthogonal alignment matrix $O_V$, defined in
	Equation~\eqref{eq:theory-mds-alignment}.
	The Nystr\"om extension ensures that the held-out observations
	are represented in the same coordinate system as the training
	observations used to fit the generators, so both sets of
	coordinates share the same alignment.
	When the roles of $\cD_1$ and $\cD_2$ are exchanged, the same
	procedure is applied using $\cD_1$ as the training sample.

	\subsection{Conditional generators in embedded spaces}\label{sec:method-generators}
	
	For $S\in\{X,Y\}$, we approximate the conditional distribution of
	$\phi_S(S)$ given $\phi_Z(Z)$ using a generative neural network
	(generator). Conditioning on $\phi_Z(Z)$ rather than $Z$
	(or equivalently $\Phi_Z(Z)$) discards the coordinates
	$[\Phi_Z(Z)]_j$ for $j>p_Z$, introducing a truncation error
	$B_S$ defined in Equation~\eqref{eq:theory-conditioning-truncation}.
	This error vanishes if the discarded coordinates carry no additional
	information about $\phi_S(S)$ given $\phi_Z(Z)$.
	Under Assumption~\ref{ass:theory-spectral-truncation}, $B_S$
	decreases polynomially or exponentially in $p_Z$, depending on
	the eigenvalue decay of the embedding of $Z$.
	
	Let $U\sim\bP_U$ be a $p_N$-dimensional noise vector independent
	of the data, where $\bP_U$ is a continuous distribution such as
	the standard Gaussian or uniform distribution on $[-1,1]^{p_N}$.
	The dimension $p_N$ can be any positive integer, and we use
	$p_N=30$ in all simulations.
	By the noise-outsourcing lemma
	\citep[Lemma~3.22]{kallenberg2002foundations},
	there exists a measurable oracle generator
	$g_{S,0}:\R^{p_Z}\times\R^{p_N}\to\R^{p_S}$ such that
	$(g_{S,0}\{\phi_Z(Z),U\},\phi_Z(Z))\stackrel{d}{=}(\phi_S(S),\phi_Z(Z))$.
	Thus, for almost every $w$ under the distribution of $\phi_Z(Z)$,
	$g_{S,0}(w,U)$ follows the conditional distribution of
	$\phi_S(S)$ given $\phi_Z(Z)=w$, and independent draws of $U$
	provide samples from this distribution.
	The oracle generator need not be unique. In
	Section~\ref{sec:theory-lowdim}, we assume that it can be
	chosen to have a smooth compositional structure.
	
	As discussed in Section~\ref{sec:method-mds}, the estimated
	coordinates of $\phi_Z(Z_i)$ and $\phi_S(S_i)$ are identifiable
	only up to orthogonal transformations. Consequently, a generator
	fitted to the pairs
	$\{\wh\phi_Z(Z_i),\wh\phi_S(S_i)\}_{i\in\cJ_2}$
	targets the rotated oracle generator
	$(w,u)\mapsto O_S^{\top}g_{S,0}(O_Zw,u)$,
	rather than $g_{S,0}$ itself, where $O_Z$ and $O_S$ are
	data-dependent orthogonal matrices determined by $\cD_2$.
	We therefore consider a class $\cG_{0,S}$ of feedforward
	neural networks $g:\R^{p_Z}\times\R^{p_N}\to\R^{p_S}$
	to approximate the oracle generator up to these transformations.

	We train the generators by empirical risk minimization using the
	kernel loss of engression \citep[Section~G.1]{shen2025engression}.
	Let $k_S^{\rm e}:\R^{p_S}\times\R^{p_S}\to\R$ be a bounded,
	continuous, characteristic kernel
	\citep[Section~2.2]{fukumizu2007kernel}, such as the
	Gaussian kernel. Choose an integer $M_{\rm tr}\ge2$ and
	draw independent noise vectors $U_{S,ia}\sim\bP_U$,
	for $i\in\cJ_2$ and $a\in[M_{\rm tr}]$, independently
	of the observations. For $g\in\cG_{0,S}$, write
	$G_{S,ia}(g)=g\{\wh\phi_Z(Z_i),U_{S,ia}\}$ and define
	the empirical risk
	\begin{equation}\label{eq:method-sampled-loss}
		\begin{aligned}
			\wh L_{S,n}(g)=\frac1n
			\sum_{i\in\cJ_2}\Biggl[
			&\frac1{M_{\rm tr}(M_{\rm tr}-1)}
			\sum_{\substack{a,a'\in[M_{\rm tr}]\\a\ne a'}}
			k_S^{\rm e}\{G_{S,ia}(g),G_{S,ia'}(g)\}\\
			&-\frac2{M_{\rm tr}}\sum_{a=1}^{M_{\rm tr}}
			k_S^{\rm e}\{\wh\phi_S(S_i),G_{S,ia}(g)\}
			\Biggr].
		\end{aligned}
	\end{equation}
	Let $\wh g_S$ denote an approximate minimizer of
	$\wh L_{S,n}$ over $\cG_{0,S}$.
	To understand this risk, consider its population counterpart
	with the true embedding coordinates. Up to a term independent
	of $g$, it equals the expected squared maximum mean discrepancy
	\citep[MMD;][]{gretton2012kernel} between the conditional
	distributions of $g\{\phi_Z(Z),U\}$ and $\phi_S(S)$
	given $\phi_Z(Z)$. Since $k_S^{\rm e}$ is characteristic,
	this discrepancy vanishes if and only if the two conditional
	distributions agree almost surely. Thus, minimizing the empirical
	risk $\wh L_{S,n}$ aims to learn the conditional distribution
	of the embedded response from the estimated coordinates.

	\subsection{Test statistic and bootstrap calibration}
	\label{sec:method-statistic}\label{sec:method-bootstrap}
	
	We extend the maximum mean discrepancy measure of conditional
	independence (MMDCI) of \citet{zhang2024doubly} to random objects
	and construct a test statistic using the estimated embeddings and
	conditional generators. For each $V\in\{X,Y,Z\}$, consider a radial
	kernel on $\ell^2$ of the form
	\begin{equation}\label{eq:method-full-kernels}
		k_{V,\infty}(a,a')=f_V(\|a-a'\|_{\ell^2}^2),
		\qquad a,a'\in\ell^2,
	\end{equation}
	where $f_V:[0,\infty)\to\R$. For example,
	$f_V(r)=\exp(-r/h_V)$ with bandwidth $h_V>0$ gives the
	Gaussian kernel used in Section~\ref{sec:simulation}.
	Let $\cH_V$ denote the corresponding RKHS, with feature map
	$\psi_{V,\infty}(a)=k_{V,\infty}(a,\cdot)$.
	For $z,z'\in\sZ$, define $k_Z(z,z')=f_Z\{d_Z(z,z')\}$,
	which equals $k_{Z,\infty}\{\Phi_Z(z),\Phi_Z(z')\}$
	for $z,z'\in\supp(\bP_Z)$.
	We then define the product kernel on $\sX\times\sY\times\sZ$ by
	\begin{equation}\label{eq:method-population-product-kernel}
		k_0\{(x,y,z),(x',y',z')\}
		=f_X\{d_X(x,x')\}f_Y\{d_Y(y,y')\}k_Z(z,z'),
	\end{equation}
	and denote its RKHS by $\cH_0$.
	
	Let $\bP_{XYZ}$ be the joint distribution of $(X,Y,Z)$,
	and let $\bP_{XYZ}^0$ be the corresponding distribution under
	conditional independence, defined by
	$\bP_{XYZ}^0(dx,dy,dz)=
	\bP_{X\mid Z}(dx\mid z)\bP_{Y\mid Z}(dy\mid z)\bP_Z(dz)$.
	Thus, $\bP_{XYZ}^0$ preserves the conditional distributions of
	$X$ and $Y$ given $Z$ while imposing conditional independence.
	We define MMDCI as the squared maximum mean discrepancy between
	$\bP_{XYZ}$ and $\bP_{XYZ}^0$:
	\begin{equation}\label{eq:method-population-mmdci}
		\MMDCI(\bP_{XYZ})=
		\left\|\int k_0\{(x,y,z),\cdot\}\,
		d(\bP_{XYZ}-\bP_{XYZ}^0)(x,y,z)\right\|_{\cH_0}^2.
	\end{equation}
	When $k_0$ is characteristic, as is the case for the kernels
	considered in Section~\ref{sec:theory-generator-null}, including
	Gaussian kernels, $\MMDCI(\bP_{XYZ})=0$ if and only if
	$\bP_{XYZ}=\bP_{XYZ}^0$, or equivalently,
	$X\indep Y\mid Z$. This motivates rejecting $H_0$ for
	large values of an estimate of $\MMDCI(\bP_{XYZ})$.
	
	To construct an estimator, we use an equivalent representation
	of MMDCI in terms of conditional mean embeddings.
	For $S\in\{X,Y\}$, define
	$\mu_{S,\infty}(z)=
	\E[\psi_{S,\infty}\{\Phi_S(S)\}\mid Z=z]\in\cH_S$
	and the residual kernel
	\begin{equation}\label{eq:method-population-residual-kernel}
		U_S^0(s,s';z,z')=
		\left\li\psi_{S,\infty}\{\Phi_S(s)\}-\mu_{S,\infty}(z),
		\psi_{S,\infty}\{\Phi_S(s')\}-\mu_{S,\infty}(z')
		\right\ri_{\cH_S}.
	\end{equation}
	By \citet{zhang2024doubly},
	\begin{equation}\label{eq:method-population-residual-mmdci}
		\MMDCI(\bP_{XYZ})=
		\E\bigl[U_X^0(X,X';Z,Z')U_Y^0(Y,Y';Z,Z')k_Z(Z,Z')\bigr],
	\end{equation}
	where $(X',Y',Z')$ is an independent copy of $(X,Y,Z)$.
	An important property of this representation is double robustness:
	under $H_0$, the right-hand side remains zero if either
	$\mu_{X,\infty}$ or $\mu_{Y,\infty}$ is replaced by an arbitrary
	bounded measurable function taking values in the corresponding RKHS
	\citep[Definition~1]{zhang2024doubly}.
	At the sample level, this allows relatively slow convergence of
	the estimated conditional mean embeddings, provided that the product
	of their estimation errors is sufficiently small; see
	condition~(\ref{eq:theory-null-generator-rates}) in
	Theorem~\ref{cor:paper-null}.
	
	We estimate the conditional mean embeddings using samples from
	the fitted generators. For $S\in\{X,Y\}$, first define a kernel
	on the truncated embedding space $\R^{p_S}$ by
	\begin{equation}\label{eq:method-response-kernel}
		k_S(x,x')=k_{S,\infty}(J_{p_S}x,J_{p_S}x')
		=f_S(\|x-x'\|_2^2),\qquad x,x'\in\R^{p_S},
	\end{equation}
	and let $\cH_{S,p_S}$ be its RKHS, with feature map
	$\psi_S(x)=k_S(x,\cdot)$.
	We approximate $\mu_{S,\infty}(z)$ using the conditional mean
	embedding of $\phi_S(S)$ given $\phi_Z(Z)=\phi_Z(z)$,
	which is then estimated by Monte Carlo samples from $\wh g_S$.
	Specifically, for each $i\in\cJ_1$, draw $M_{\rm ev}$
	independent noise vectors $U_{S,i,m}\sim\bP_U$ and generate
	$\wh S_{i,m}=\wh g_S(\wt Z_i,U_{S,i,m})$ for $m\in[M_{\rm ev}]$,
	where the noise vectors are mutually independent across $S,i,m$
	and independent of the observations. The sample average
	$M_{\rm ev}^{-1}\sum_{m=1}^{M_{\rm ev}}\psi_S(\wh S_{i,m})$
	estimates the conditional mean embedding in the truncated space.
	Substituting these estimates into the residual kernel gives
	\begin{equation}\label{eq:method-residual-kernel}
		\begin{aligned}
			\wh U_{S,ij}={}&
			k_S(\wt S_i,\wt S_j)
			-\frac1{M_{\rm ev}}\sum_{m=1}^{M_{\rm ev}}
			\bigl\{k_S(\wt S_i,\wh S_{j,m})
			+k_S(\wh S_{i,m},\wt S_j)\bigr\}\\
			&+\frac1{M_{\rm ev}^2}
			\sum_{m=1}^{M_{\rm ev}}\sum_{\ell=1}^{M_{\rm ev}}
			k_S(\wh S_{i,m},\wh S_{j,\ell}).
		\end{aligned}
	\end{equation}
	Using the estimated residual kernels, we construct the
	cross-fitted statistic on $\cD_1$ as
	\begin{equation}\label{eq:method-fold-statistic}
		\wh T_1=\frac1{n(n-1)}
		\sum_{\substack{i,j\in\cJ_1\\i\ne j}}
		\wh U_{X,ij}\wh U_{Y,ij}k_Z(Z_i,Z_j).
	\end{equation}
	Exchanging the roles of $\cD_1$ and $\cD_2$ gives
	$\wh T_2$, and the final test statistic is $\wh T_n=\wh T_1+\wh T_2$.
	
	Under $H_0$, $n\wh T_n$ converges in distribution to a weighted
	sum of centered chi-squared random variables, with weights
	depending on the unknown distribution of $(X,Y,Z)$
	(Remark~\ref{rem:theory-full-statistic}).
	We therefore use a wild bootstrap adapted from
	\citet{zhang2024doubly} to approximate the null distribution.
	For $B\ge1$ bootstrap repetitions, draw
	$e_i^{(b)}\stackrel{\rm iid}{\sim}N(0,1)$ for
	$i\in[2n]$ and $b\in[B]$, independently of all previous
	randomness. For each repetition $b$, compute
	\begin{equation}\label{eq:method-bootstrap-statistic}
		\wh T_n^{*(b)}=\wh T_1^{*(b)}+\wh T_2^{*(b)},
		\qquad
		\wh T_1^{*(b)}=\frac1{n(n-1)}
		\sum_{\substack{i,j\in\cJ_1\\i\ne j}}
		e_i^{(b)}e_j^{(b)}
		\wh U_{X,ij}\wh U_{Y,ij}k_Z(Z_i,Z_j),
	\end{equation}
	where $\wh T_2^{*(b)}$ is defined analogously using $\cD_2$.
	The bootstrap $p$-value is
	\begin{equation}\label{eq:method-bootstrap-pvalue}
		\wh p=\frac{1+\sum_{b=1}^B
			\indic\{\wh T_n^{*(b)}\ge\wh T_n\}}{B+1}.
	\end{equation}
	We reject $H_0$ at nominal level $\alpha\in(0,1)$
	when $\wh p\le\alpha$.
	
	\medskip
	\noindent{\bf Euclidean conditioning variables.}
	\label{sec:method-vector-z}
	When $Z\in\R^{p_Z}$ is a high-dimensional random vector,
	we can use $Z$ directly as input to the generators, avoiding
	the estimation and truncation of its Hilbert space embedding.
	In this case, MDS is applied only to $X$ and $Y$.
	We replace $\wh\phi_Z(Z_i)$ in
	Equation~\eqref{eq:method-sampled-loss} and $\wt Z_i$ in
	the definition of $\wh S_{i,m}$ with $Z_i$.
	The generated samples become
	$\wh S_{i,m}=\wh g_S(Z_i,U_{S,i,m})$,
	and the test statistic and bootstrap procedure are otherwise
	unchanged.

	\section{Theoretical Guarantees and Error Analysis}\label{sec:theory}
	
	We first establish general conditions for the validity and power of the
	proposed test, and then verify these conditions through bounds on the
	MDS embedding and generator estimation errors. We focus on $\wh T_1$;
	results for $\wh T_2$ follow by exchanging the two data splits.
	Section~\ref{sec:theory-mean-error} decomposes and bounds the errors
	of the estimated conditional mean embeddings.
	Section~\ref{sec:theory-generator-null} uses these bounds to establish
	the asymptotic null distribution, bootstrap validity, and power of
	the test, including the case of Euclidean conditioning variables.
	Section~\ref{sec:theory-mds} bounds the MDS and Nystr\"om extension errors,
	while Section~\ref{sec:theory-lowdim} develops approximation and
	complexity bounds for neural generators with low-dimensional structure.
	Finally, Section~\ref{sec:theory-growing-rates} combines these results
	to derive explicit convergence rates with growing embedding dimensions.
	All proofs are provided in the appendix.
	
	Throughout this section, we use the same kernel for generator
	training and test statistic construction, namely $k_S^{\rm e}=k_S$.
	We consider the radial kernels in Equation~\eqref{eq:method-full-kernels}
	with $f_V(r)=\int_{[0,\infty)}e^{-ur}\,d\nu_V(u)$, where $\nu_V$
	is a finite measure on $[0,\infty)$ satisfying
	$\nu_V((0,\infty))>0$ and
	$\int_{[0,\infty)}u\,d\nu_V(u)<\infty$.
	This class includes the Gaussian kernel, obtained by taking
	$\nu_V$ to be a point mass at $1/h_V$.
	By \citet[Theorem~3.1]{ziegel2024characteristic}, the resulting
	product kernel $k_0$ in
	Equation~\eqref{eq:method-population-product-kernel} is characteristic.
	The condition $\nu_V((0,\infty))>0$ excludes constant kernels,
	while the finite first moment ensures that the feature map
	$\psi_{V,\infty}$ is Lipschitz.
	Throughout, $C$ denotes a positive constant that may depend
	on the fixed kernels but not on $n$, and may change from line to line.
	
	Let $\cF_2$ denote the sigma-algebra generated by $\cD_2$
	and the random variables used to train $\wh g_X$ and $\wh g_Y$.
	Let $(X^\circ,Y^\circ,Z^\circ)$ be an independent copy of
	$(X,Y,Z)$, independent of $\cF_2$, representing a new observation
	from $\cD_1$. For object-valued conditioning variables, we assume
	that $\phi_Z(Z^\circ)$, $\wh\phi_Z(Z^\circ)$, and
	$\wh\phi_Z(Z_i)$ for $i\in\cJ_2$ all lie almost surely in
	$\cD_Z=\{w\in\R^{p_Z}:\|w\|\le R_0\}$ for some constant $R_0\ge1$.

	\subsection{Error bounds for estimated conditional mean embeddings}\label{sec:theory-mean-error}
	
	The fitted generators estimate the conditional mean embeddings used in
	the residual kernels of the test statistic. Their estimation errors
	arise from constructing and truncating the Hilbert embeddings and
	from learning the conditional distributions. We first define these
	errors and then bound their combined effect in
	Theorem~\ref{thm:paper-generator}.
	
	We first quantify the MDS embedding and Nystr\"om extension errors.
	Since the MDS coordinates are identifiable only up to orthogonal
	transformations, we align the estimated coordinates with their
	population counterparts. For each $V\in\{X,Y,Z\}$, let $O_V$ be
	a random orthogonal matrix depending only on the training observations
	$\{V_i:i\in\cJ_2\}$, chosen to minimize the average squared
	alignment error:
	\begin{equation}\label{eq:theory-mds-alignment}
		O_V\in\argmin_{\substack{Q\in\R^{p_V\times p_V}\\Q^{\top}Q=I_{p_V}}}
		\frac1n\sum_{i\in\cJ_2}
		\|Q\wh\phi_V(V_i)-\phi_V(V_i)\|^2.
	\end{equation}
	For an independent copy $V^\circ$ of $V$, define the combined
	training and out-of-sample coordinate error by
	\begin{equation}\label{eq:theory-mds-error}
		\rho_{V,n}^2=\E\left[
		\frac1n\sum_{i\in\cJ_2}\|O_V\wh\phi_V(V_i)-\phi_V(V_i)\|^2
		+\|O_V\wh\phi_V(V^\circ)-\phi_V(V^\circ)\|^2
		\right].
	\end{equation}
	Equation~\eqref{eq:theory-mds-error} uses the same alignment $O_V$ for two types of errors: the average
	error of the MDS embeddings $\{\wh\phi_V(V_i):i\in\cJ_2\}$, to which the
	generators are fitted, and the error of the Nystr\"om extension at an
	independent object $V^\circ$, which gives the embedded data used in $\wh T_1$. Theorem~\ref{thm:paper-mds} in
	Section~\ref{sec:theory-mds} gives an explicit bound on
	$\rho_{V,n}^2$.
	
	For the response $S\in\{X,Y\}$, the truncated oracle embedding $\phi_S(S)$ omits the coordinates $j>p_S$ of $\Phi_S(S)$,
	whose expected squared norm is the spectral tail $\sum_{j>p_S}\theta_{S,j}$.
	For the conditioning variable, the oracle generators $g_{S,0}$ condition on $\phi_Z(Z)$
	instead of $\Phi_Z(Z)$, and we quantify the induced error by 	\begin{equation}\label{eq:theory-conditioning-truncation}
		B_S=\E\left\|
		\E[\psi_S\{\phi_S(S)\}\mid Z]
		-\E[\psi_S\{\phi_S(S)\}\mid\phi_Z(Z)]\right\|_{\cH_{S,p_S}}^2.
	\end{equation}
	Together, $B_S$ and $\sum_{j>p_S}\theta_{S,j}$ quantify the error from truncating the infinite-dimensional oracle embedding to a $p_V$-dimensional vector.
	
	We next turn to the generator estimation error, which consists of approximation error, stochastic error and optimization error. As explained in Section~\ref{sec:method-generators}, the fitted generator targets
	$(w,u)\mapsto O_S^{\top}g_{S,0}(O_Zw,u)$. Since $O_Z$ and $O_S$ are unknown, we need to quantify the approximation error of $(w,u)\mapsto Q_S^{\top}g_{S,0}(Q_Zw,u)$ for all
	orthogonal matrices $Q_Z\in\R^{p_Z\times p_Z}$ and $Q_S\in\R^{p_S\times p_S}$ on a larger class of functions than the class of networks $\cG_{0,S}$ in Section~\ref{sec:method-generators}.
	For any pair $Q=(Q_S,Q_Z)$ of orthogonal matrices of sizes $p_S$ and $p_Z$, let
	\begin{equation}\label{eq:theory-rotation-classes}
		\cG_{S,Q}=\{(w,u)\mapsto Q_Sg(Q_Z^{\top}w,u):g\in\cG_{0,S}\},
	\end{equation}
	and we define the enlarged function class as $\cG_S=\bigcup_Q\cG_{S,Q}$.
	Define the approximation error
	\begin{equation}\label{eq:theory-approximation-error}
		A_S=\E\left[
		\inf_{g\in\cG_{S,(O_S,O_Z)}}
		\int\|\E_U\psi_S[g\{\phi_Z(z),U\}]-\E_U\psi_S[g_{S,0}\{\phi_Z(z),U\}]\|_{\cH_{S,p_S}}^2
		\,d\bP_Z(z)\right].
	\end{equation}
	Here $A_S$ quantifies how well the network class
	$\cG_{S,(O_S,O_Z)}$ approximates the conditional distribution of
	$\phi_S(S)$ given $\phi_Z(Z)$.
	The integral in
	Equation~\eqref{eq:theory-approximation-error} for a fixed $g$ equals, up to
	a term that does not depend on $g$, the expectation of the loss in
	Equation~\eqref{eq:method-sampled-loss} when $\{\wh\phi_Z(Z_i),\wh\phi_S(S_i)\}$ are
	replaced by the oracle embeddings $\{\phi_Z(Z_i),\phi_S(S_i)\}$. Thus $A_S$
	is the expected smallest excess loss attainable in the random class
	$\cG_{S,(O_S,O_Z)}$.
	
	The stochastic error of the fitted generator depends on the complexity of
	the generator class. In the literature, this error is typically bounded
	through the covering numbers of the class; see Theorem~11.4 in
	\citet{gyorfi2002distribution} and Lemma~4 in
	\citet{schmidthieber2020nonparametric}. We therefore adopt the following
	standard assumption on the covering numbers, imposed on the enlarged class
	$\cG_S$.
	For measurable maps $g,h:\R^{p_Z}\times\R^{p_N}\to\R^{p_S}$, set
	\begin{equation}\label{eq:theory-generator-metric}
		d_{\cG_S}(g,h)=
		\sup_{\substack{w\in\cD_Z\\u\in\supp(\bP_U)}}\|g(w,u)-h(w,u)\|,
	\end{equation}
	which is the distance on $\cG_S$.
	We use covering numbers whose centers need not belong to the covered set
	\citep[Section~6.9]{rauhut2010compressive}. In particular,
$\cN(\varepsilon,\cG_S,d_{\cG_S})$ is the minimum number of measurable maps
	such that every $g\in\cG_S$ is within $\varepsilon$ under $d_{\cG_S}$
	of at least one of these maps.
	Besides the covering numbers, Assumption~\ref{ass:theory-generator-class}
	below contains a Lipschitz condition, which controls how the MDS embedding/Nystr\"om extension
	error of $Z$ propagates through the generator since the generators are fitted and evaluated at the
	estimated embeddings $\wh \phi_Z$ rather than at $\phi_Z$.
	
	\begin{assumption}\label{ass:theory-generator-class}
		For $L_{\mu,S}\ge0$, $v_S\ge1$, and $\Lambda_S\ge e$,
	\begin{compactenum}[(i)]
		\item\label{eq:theory-mean-lipschitz}
		$\|\E_U\psi_S\{g(w,U)\}-\E_U\psi_S\{g(w',U)\}\|_{\cH_{S,p_S}}
		\le L_{\mu,S}\|w-w'\|$ for $g\in\cG_S$ and $w,w'\in\cD_Z$;
		\item\label{eq:theory-generator-covering}
		$\log\cN(\varepsilon,\cG_S,d_{\cG_S})\le v_S\log(\Lambda_S/\varepsilon)$
		for $0<\varepsilon\le1$.
		\item\label{eq:theory-optimization}
		For an optimization error $\delta_S\ge0$, the fitted generator
		$\wh g_S\in\cG_{0,S}$ satisfies
		$\wh L_{S,n}(\wh g_S)\le\inf_{g\in\cG_{0,S}}\wh L_{S,n}(g)+\delta_S$
		almost surely.
	\end{compactenum}
\end{assumption}
	
	By Jensen's inequality and the Lipschitz property of $\psi_S$,
	Assumption~\ref{ass:theory-generator-class}(\ref{eq:theory-mean-lipschitz}) holds when the networks are
	uniformly Lipschitz in $w$, as in
	Proposition~\ref{prop:paper-network-covering} in
	Appendix~\ref{app:proof-network-covering}.
	Assumption~\ref{ass:theory-generator-class}(\ref{eq:theory-generator-covering}) is a parametric entropy bound
	in which $v_S$ plays the role of the number of parameters; bounds of this
	form hold for sparse ReLU networks
	\citep[Lemma~5]{schmidthieber2020nonparametric}; see also the correction
	by \citet{schmidtvu2024correction}.
	
	Define the stochastic error terms
	\begin{equation}\label{eq:theory-stochastic-rates}
		\tau_S^2=\frac{2v_S\log(\Lambda_S n)}n,\qquad
		J_S=\sqrt{\frac{v_S\log(\Lambda_S nM_{\rm tr})}{nM_{\rm tr}}}
		+\frac1{nM_{\rm tr}},
	\end{equation}
	where $\tau_S^2$ is the statistical error from fitting the generator to $n$
	observations and $J_S$ is the Monte Carlo error from the $M_{\rm tr}$ noise
	draws per observation in Equation~\eqref{eq:method-sampled-loss}.
	The combined error is
	\begin{equation}\label{eq:theory-error-budget}
		\cE_S=A_S+\tau_S^2+J_S+\delta_S
		+\rho_{S,n}^2+B_S+L_{\mu,S}^2\rho_{Z,n}^2.
	\end{equation}
	The generator estimation error $A_S+\tau_S^2+J_S+\delta_S$ has the same
	four components as in Theorem~D.3 of \citet{chai2026neural} for Euclidean data.
	The remaining terms $\rho_{S,n}^2+B_S+L_{\mu,S}^2\rho_{Z,n}^2$ form the MDS
	embedding/Nystr\"om extension error: $\rho_{S,n}^2$ and
	$L_{\mu,S}^2\rho_{Z,n}^2$ arise from estimating $\phi_S$ and $\phi_Z$ by MDS,
	and $B_S$ from truncating $\Phi_Z(Z)$. Importantly,
	a larger network class can reduce $A_S$, but it may also increase the
	Lipschitz constant $L_{\mu,S}$, amplifying the effect of estimating
	$\phi_Z$. Proposition~\ref{prop:paper-lowdim-approximation} in
	Section~\ref{sec:theory-lowdim} shows how to choose the network class
	so that increasing its width improves the bound on $A_S$ while keeping
	the bound on $L_{\mu,S}$ independent of the width.
	
	To state the bound, for $S\in\{X,Y\}$, define
	\begin{equation}\label{eq:theory-conditional-mean-error}
		\begin{aligned}
			\wh m_S(z)
			&=\E_U\psi_{S,\infty}\!\left[
			J_{p_S}O_S\wh g_S\{\wh\phi_Z(z),U\}\right],\ \ 
			R_S^2
			=\E\|\wh m_S(Z^\circ)-\mu_{S,\infty}(Z^\circ)\|_{\cH_S}^2.
		\end{aligned}
	\end{equation}
	Here $J_{p_S}O_S$ rotates the generated vectors to the orientation of the truncated oracle embedding
	$\phi_S$ and projects them into $\ell^2$. Thus $R_S^2$ is the mean squared error
	of $\wh m_S(Z^\circ)$, the conditional mean embedding of the generated vectors given
	$\wh\phi_Z(Z^\circ)$, as an estimate of the conditional mean embedding
	$\mu_{S,\infty}(Z^\circ)$ in Equation~\eqref{eq:method-population-residual-kernel}.
	It is the main estimation error described at the beginning of this
	subsection, without the Monte Carlo error from the $M_{\rm ev}$ draws, which
	vanishes as $M_{\rm ev}\to\infty$.
	The following theorem gives a bound on the conditional mean embedding estimation error $R_S^2$.
	\begin{theorem}\label{thm:paper-generator}
		Under Assumption~\ref{ass:theory-generator-class}, for each
		$S\in\{X,Y\}$, the mean squared error $R_S^2$ of the estimated
		conditional mean embedding satisfies
	$R_S^2\le C\{\cE_S+\sum_{j>p_S}\theta_{S,j}\}$.
\end{theorem}
	
	The spectral tail $\sum_{j>p_S}\theta_{S,j}$ appears in
the bound of Theorem~\ref{thm:paper-generator} because $R_S^2$ measures the
	error relative to $\mu_{S,\infty}$, the conditional mean embedding of the
	full $\Phi_S(S)$, whereas the generators model only the truncated
	$\phi_S(S)$.
	
	\subsection{Asymptotic validity and power of the proposed test}\label{sec:theory-generator-null}
	
	We now use the preceding error bound to study the asymptotic properties
	of the test. To derive the limiting null distribution of $n\wh T_1$, note that the oracle version of $\wh T_1$ replaces the estimated kernels $\wh U_{S,ij}$
	in Equation~\eqref{eq:method-fold-statistic} by the population kernels
	$U_S^0(S_i,S_j;Z_i,Z_j)$ in Equation~\eqref{eq:method-population-residual-kernel}.
	It can be written as
	$T_1=\frac1{n(n-1)}\sum_{\substack{i,j\in\cJ_1\\i\ne j}}\li W_i^0,W_j^0\ri$,
	where $W_i^0$ is
	\begin{equation}\label{eq:theory-population-tensor}
		W^0=\bigl[\psi_{X,\infty}\{\Phi_X(X)\}-\mu_{X,\infty}(Z)\bigr]
		\otimes\bigl[\psi_{Y,\infty}\{\Phi_Y(Y)\}-\mu_{Y,\infty}(Z)\bigr]\otimes\psi_{Z,\infty}\{\Phi_Z(Z)\}
	\end{equation}
	evaluated at $(X_i,Y_i,Z_i)$. Under $H_0$, $X$ and $Y$ are conditionally
	independent given $Z$, so $\E W^0=0$ and $T_1$ is a degenerate U-statistic.
	Denote the covariance operator of $W^0$ by $\Gamma\neq 0$, and let $(\lambda_j)_{j=1}^\infty$
	be its nonnegative eigenvalues.
	The following theorem gives the limiting null distribution of $n\wh T_1$
	under conditions on the combined errors $\cE_S$, the number $M_{\rm ev}$ of samples from the trained generator,
	and the spectral tails.
	
	\begin{theorem}\label{cor:paper-null}
		Suppose that $H_0$ and Assumption~\ref{ass:theory-generator-class} hold, and
		that as $n\to\infty$,
		{\setcounter{enumi}{0}\renewcommand{\theenumi}{\roman{enumi}}%
			\refstepcounter{enumi}(i)\label{eq:theory-null-generator-rates}
			$\cE_X+\cE_Y+n\cE_X\cE_Y\to0$;
			\refstepcounter{enumi}(ii)\label{eq:theory-null-monte-carlo}
			$M_{\rm ev}\to\infty$; and
			\refstepcounter{enumi}(iii)\label{eq:theory-null-feature-rates}
			$\sum_{S\in\{X,Y\}}\sum_{j>p_S}\theta_{S,j}\to0$.}
		Then $n(\wh T_1-T_1)\stackrel{p}{\to}0$ and
	$n\wh T_1\stackrel{d}{\to}\sum_{j\ge1}\lambda_j(N_j^2-1)$,
	where $\{N_j, j\geq 1\}$ are independent standard normal random variables.
	\end{theorem}
	
	Under conditions~(\ref{eq:theory-null-generator-rates})--(\ref{eq:theory-null-feature-rates}), the difference between $\wh T_1$ and its oracle version $T_1$ is negligible, and by the double robustness property described in Section~\ref{sec:method-statistic},
	condition~(\ref{eq:theory-null-generator-rates}) allows $\cE_X$ and $\cE_Y$
	to converge to zero at a slower rate, provided $\cE_X\cE_Y=o(n^{-1})$.
	For example, condition~(\ref{eq:theory-null-generator-rates}) holds if
	$\cE_S=o(n^{-1/2})$ for $S\in\{X,Y\}$, which is proved in
	Section~\ref{sec:theory-growing-rates} under smoothness assumptions.
	Similar product conditions appear in conditional independence tests for random
	vectors, such as the generalized covariance measure
	\citep[Theorem~6]{ShahPeters2020} and the doubly robust test of
	\citet[Assumption~2]{zhang2024doubly}.
	As in \citet{zhang2024doubly}, condition~(\ref{eq:theory-null-monte-carlo}) allows
	$M_{\rm ev}$ to diverge at any rate. Condition~(\ref{eq:theory-null-feature-rates})
	requires only that the response spectral tails $\sum_{j>p_S}\theta_{S,j}$ tend
	to zero, which holds if $p_S\to\infty$.
	
	\begin{remark}\label{rem:theory-full-statistic}
		The limiting null distribution in Theorem~\ref{cor:paper-null}
		is the typical limiting distribution of degenerate U-statistics of order 2
		\citep[Section~5.5.2]{serfling1980approximation}.
		Since $T_1$ and the oracle version of $\wh T_2$ are independent and
		identically distributed, under the conditions of Theorem~\ref{cor:paper-null},
	$n\wh T_n\stackrel{d}{\to}\sum_{j\ge1}\lambda_j(\chi_{2,j}^2-2)$,
	where the $\chi_{2,j}^2$ are independent chi-squared random variables with two
		degrees of freedom. The weights $\lambda_j$ depend on the unknown distribution
		of $(X,Y,Z)$, so we approximate the null distribution of $\wh T_n$ by the wild
		bootstrap in Section~\ref{sec:method-bootstrap}.
	\end{remark}
	
	Next, we study the rejection probability under the null and under
	alternatives, including local alternatives whose distribution $\bP_{XYZ}$
	changes with $n$. We measure the departure of $\bP_{XYZ}$ from $H_0$ by
	\begin{equation}\label{eq:theory-alternative-signal}
		s_n=\{\MMDCI(\bP_{XYZ})\}^{1/2}=\|\E W^0\|.
	\end{equation}
	Under local alternatives, we let $s_n$ decay to zero as $n$ grows.
	Such alternatives show how small a departure from $H_0$ the test can detect:
	$\wh T_1$ estimates $\MMDCI(\bP_{XYZ})=s_n^2$, while $n\wh T_1$ under
	$H_0$ is of order $O_p(1)$ by Theorem~\ref{cor:paper-null}, so the test can be
	expected to detect $s_n$ larger than $n^{-1/2}$. Theorem~\ref{thm:paper-power} below makes this precise.
	As in Theorem~\ref{cor:paper-null}, the asymptotic properties of $n\wh T_1$ are
	mainly determined by the eigenvalues of the covariance operator of $W^0$, now under a distribution $\bP_{XYZ}$ that may change with $n$. Let
	$\Gamma_n$ be this operator, with eigenvalues
	$\lambda_{n,1}\ge\lambda_{n,2}\ge\cdots\ge0$. The following assumption gives a uniform lower
	bound on its squared Hilbert--Schmidt norm.
	\begin{assumption}\label{ass:theory-vector-variance}
		\sloppy
		There is a constant $\sigma_0>0$ such that
		$\inf_{n\ge2}2\sum_{j\ge1}\lambda_{n,j}^2=\inf_{n\ge2}2\|\Gamma_n\|_{\mathrm{HS}}^2\ge\sigma_0^2$.
	\end{assumption}
	Since $2\sum_{j\ge1}\lambda_{n,j}^2$ is the variance of
	$\sum_{j\ge1}\lambda_{n,j}(N_j^2-1)$, Assumption~\ref{ass:theory-vector-variance}
	ensures that the weighted chi-squared distributions approximating the null
	behavior of $n\wh T_1$ do not degenerate. For a fixed $\bP_{XYZ}$, it holds
	whenever $\Gamma_n=\Gamma\ne0$.
	
	Define $q_{S,n}^2=\rho_{S,n}^2+\sum_{j>p_S}\theta_{S,j}$ for $S\in\{X,Y\}$ and
	\[
	\begin{aligned}
		d_{X,n}^2&=\E\|\E[\psi_{X,\infty}\{\Phi_X(X)\}\mid Y,Z]-\mu_{X,\infty}(Z)\|^2\\
		&\quad+\E\|\E[\psi_X\{\phi_X(X)\}\mid Y,Z]-\E[\psi_X\{\phi_X(X)\}\mid Z]\|^2,
	\end{aligned}
	\]
	and define $d_{Y,n}^2$ by exchanging $X$ and $Y$. Here $q_{S,n}^2$ combines the MDS
	embedding/\allowbreak Nystr\"om extension error and the spectral tail of $S$, the two errors in
	representing $\Phi_S(S)$ by its estimated embedding. The quantities
	$d_{X,n}^2$ and $d_{Y,n}^2$ measure the departures from $H_0$ of the conditional mean
	embeddings of the full $\Phi_S(S)$ and of the truncated $\phi_S(S)$:
	under $H_0$, $X$ and $Y$ are conditionally independent given $Z$, so
	$d_{X,n}=d_{Y,n}=0$.
	
	\begin{theorem}\label{thm:paper-power}\label{thm:paper-alternative}\label{thm:paper-bootstrap}
		Suppose that Assumptions~\ref{ass:theory-generator-class}
		and~\ref{ass:theory-vector-variance} and
		conditions~(\ref{eq:theory-null-generator-rates})--(\ref{eq:theory-null-feature-rates})
		of Theorem~\ref{cor:paper-null} hold, and $B=B_n\to\infty$. Then for any
	$\alpha\in(0,1)$,
	\begin{compactenum}[(i)]
			\item\label{item:theory-bootstrap-size}
			Under $H_0$, $\bP(\wh p\le\alpha)\to\alpha$.
			\item\label{item:theory-alternative-fixed}
			Under a fixed alternative, $\bP(\wh p\le\alpha)\to1$.
			
			\item\label{item:theory-alternative-local}
			Suppose $s_n=n^{-a}$ for some $0<a<1/2$ and
		$n^{2a}(q_{X,n}^2d_{Y,n}^2+q_{Y,n}^2d_{X,n}^2)\to0$.
		Then $\bP(\wh p\le\alpha)\to1$.
			
			\item\label{item:theory-alternative-boundary}
			Suppose $s_n=n^{-1/2}$ and
		$n(q_{X,n}^2d_{Y,n}^2+q_{Y,n}^2d_{X,n}^2)\to0$.
		Then $\bP(\wh p\le\alpha)-\bP(L_n>c_{n,\alpha})\to0$,
		where $c_{n,\alpha}$ is the $(1-\alpha)$ quantile of
			$\sum_{j\ge1}\lambda_{n,j}(\chi_{2,j}^2-2)$ and $L_n$ is a random
			variable stochastically larger than
			$\sum_{j\ge1}\lambda_{n,j}(\chi_{2,j}^2-2)$; see
			Equation~\eqref{eq:theory-boundary-noncentral-law} in
			Appendix~\ref{app:proof-paper-power} for its definition.
			Moreover, $\liminf_{n\to\infty}\bP(\wh p\le\alpha)>\alpha$.
			
			\item\label{item:theory-alternative-weak}
			Suppose $s_n=n^{-a}$ for some $a>1/2$ and
		$n(q_{X,n}^2d_{Y,n}^2+q_{Y,n}^2d_{X,n}^2)\to0$.
			Then $\bP(\wh p\le\alpha)\to\alpha$.
	\end{compactenum}
\end{theorem}
	Theorem~\ref{thm:paper-power}(\ref{item:theory-alternative-local})--(\ref{item:theory-alternative-weak})
	shows that the test has power tending to one when $s_n$ decays more slowly
	than $n^{-1/2}$, power bounded away from $\alpha$ when $s_n=n^{-1/2}$, and
	power tending to $\alpha$ when $s_n$ decays faster than $n^{-1/2}$. These
	regimes parallel Theorems~3 and~4 of \citet{zhang2024doubly}, whose local
	alternatives perturb the conditional mean embedding of $X$ given $(Y,Z)$ by
	$n^{-a}$ times a fixed function; Theorem~\ref{thm:paper-power} does not
	assume this specific form of alternative.
	The conditions in
Theorem~\ref{thm:paper-power}(\ref{item:theory-alternative-local})--(\ref{item:theory-alternative-weak})
allow the conditional-mean
	departures $d_{S,n}^2$ to decay slowly when the MDS embedding/Nystr\"om extension errors and
	response embedding truncation errors decay fast. Since $q_{X,n}\to0$ and $q_{Y,n}\to0$ under
	conditions~(\ref{eq:theory-null-generator-rates})
	and~(\ref{eq:theory-null-feature-rates}) of Theorem~\ref{cor:paper-null},
	they also hold whenever $d_{X,n}+d_{Y,n}=O(\max\{s_n,n^{-1/2}\})$.

	\medskip
	\noindent{\bf Euclidean conditioning variables.}
	We now derive the theoretical results when the conditioning variable $Z\in\R^{p_Z}$ is a high-dimensional random vector.
	In this setting, $R_0$, $\cD_Z$, the generator classes $\cG_{S,Q}$ and $\cG_S$,
	the mean embedding $\wh m_S$,
	and all terms in the error bound in Theorem~\ref{thm:paper-generator}
	are defined by replacing $\phi_Z(z)$ and $\wh\phi_Z(z)$ by $z$.
	Note that $O_Z=I_{p_Z}$, $\rho_{Z,n}=0$, $B_S=0$ and Equation~\eqref{eq:theory-error-budget} reduces to
	\begin{equation}\label{eq:theory-vector-error-budget}
		\cE_S=A_S+\tau_S^2+J_S+\delta_S+\rho_{S,n}^2.
	\end{equation}
	
	\begin{corollary}\label{cor:paper-vector-generator}
		Suppose that $Z\in\R^{p_Z}$ and
		Assumption~\ref{ass:theory-generator-class} holds.
		Then, for each $S\in\{X,Y\}$, the mean squared error $R_S^2$
		of the estimated conditional mean embedding satisfies
	$R_S^2\le C\{\cE_S+\sum_{j>p_S}\theta_{S,j}\}$,
	where $\cE_S$ is defined in
		Equation~\eqref{eq:theory-vector-error-budget}.
	\end{corollary}
	
	Let $Q_n=\sum_{j\ge1}\lambda_{n,j}(N_j^2-1)$, where $(N_j)$ are independent
	standard normal random variables. When $p_Z$ grows, the distribution of $Z$, and hence $\Gamma_n$, changes with $n$, so $n\wh T_1$ need not have a fixed limit. The following theorem gives an approximation of the null distribution of $n\wh T_1$ when $p_Z$ grows.
	
	\begin{theorem}\label{thm:paper-vector-null}
		Under $H_0$, Assumptions~\ref{ass:theory-generator-class}
		and~\ref{ass:theory-vector-variance}, and
		conditions~(\ref{eq:theory-null-generator-rates})--(\ref{eq:theory-null-feature-rates})
		of Theorem~\ref{cor:paper-null}, we have
	$\sup_{t\in\R}|\bP(n\wh T_1\le t)-\bP(Q_n\le t)|\to0$.
\end{theorem}
	
	For the size and power of our test, the results in
	Theorem~\ref{thm:paper-power} also apply when $Z$ is a high-dimensional
	random vector. Since the proof is analogous, we omit a separate statement.
	
	\subsection{Error bounds for MDS and Nystr\"om extensions}\label{sec:theory-mds}
	
	We bound the MDS embedding/Nystr\"om extension error $\rho_{V,n}^2$ in
	Equation~\eqref{eq:theory-mds-error} for the estimated MDS embedding
	$\wh\phi_V$ in Section~\ref{sec:method-mds}.
	Define $	\mu_{2,V}=\E\|\Phi_V(V)\|_{\ell^2}^2
	=\sum_{j\ge1}\theta_{V,j}$ and the eigengap $\gamma_{V,p_V}=\theta_{V,p_V}-\theta_{V,p_V+1}$.	We impose the following fourth-moment condition on $\Phi_V(V)$.
	
	\begin{assumption}\label{ass:theory-fourth-moment}
		For each $V\in\{X,Y,Z\}$, there is a constant $\mu_{4,V}<\infty$ such that
	$\E\|\Phi_V(V)\|_{\ell^2}^4\le\mu_{4,V}$.
\end{assumption}
	Since $\|\Phi_V(t)\|_{\ell^2}^2=\kappa_V(t,t)=\E d_V(t,V)-\E d_V(V,V')/2$
	for $t\in\supp(\bP_V)$, Assumption~\ref{ass:theory-fourth-moment} holds if and only
	if $\E d_V(t_0,V)^2<\infty$ for some $t_0\in\sV$.
	For example, this condition holds whenever $d_V$ is bounded. In a Hilbert space
	with $d_V(s,t)=\|s-t\|$, it is equivalent to $\E\|V\|^2<\infty$, and for a random
	univariate distribution $V$ under the $2$-Wasserstein distance, it is equivalent
	to $\E\int x^2\,dV(x)<\infty$.
	
	Recall the alignment matrix $O_V$ defined in
	Equation~\eqref{eq:theory-mds-alignment}. The following theorem
	bounds $\rho_{V,n}^2$.
	
	\begin{theorem}\label{thm:paper-mds}
		Under Assumption~\ref{ass:theory-fourth-moment}, for each
		$V\in\{X,Y,Z\}$, the mean squared error $\rho_{V,n}^2$
		of the MDS embedding and Nystr\"om extension satisfies
	$\rho_{V,n}^2\le C\mu_{2,V}\mu_{4,V}/(n\gamma_{V,p_V}^2)$,
	where $C$ is a universal constant independent of $n$.
	\end{theorem}
	
	To see where the bound comes from, note that, since
	$d_V(s,t)=\|\Phi_V(s)-\Phi_V(t)\|_{\ell^2}^2$, the entries of $\wh\phi_V(t)$ are the
	inner products of $\Phi_V(t)-n^{-1}\sum_{i\in\cJ_2}\Phi_V(V_i)$ with the $p_V$
	leading eigenvectors of the empirical covariance operator of
	$\{\Phi_V(V_i):i\in\cJ_2\}$, whereas the entries of
	$\phi_V(t)$ are the inner products of $\Phi_V(t)$ with the $p_V$ leading
	eigenvectors of the covariance operator of $\Phi_V(V)$.
	In the bound of Theorem~\ref{thm:paper-mds}, $\mu_{4,V}/n$ bounds, up to a
	constant, the mean squared difference between the empirical covariance operator
	and the covariance operator of $\Phi_V(V)$. The factor
	$\gamma_{V,p_V}^{-2}$ converts it into an error of the space spanned by the $p_V$
	leading eigenvectors of the empirical covariance operator, through the eigenspace
	perturbation argument in
	Theorem~5.1 of \citet{daviskahan1970rotation}, and $\mu_{2,V}$ converts the
	latter into an error of the embeddings.
	Only the gap between the last retained and the first discarded eigenvalue
	enters, because $O_V$ absorbs any rotation among the retained eigenvectors.
	For a fixed $p_V$, the bound is of order $n^{-1}$; when $p_V$ grows with $n$,
	it tends to zero provided $n\gamma_{V,p_V}^2\to\infty$. For example, under the
	polynomial eigenvalue decay in
	Assumption~\ref{ass:theory-spectral-truncation}(\ref{item:theory-spectral-decay}),
	$\gamma_{V,p_V}^{-2}$ grows polynomially in $p_V$ and $p_V$ may grow polynomially
	in $n$, whereas under the exponential decay, $\gamma_{V,p_V}^{-2}$ grows
	exponentially in $p_V$ and $p_V$ may grow only logarithmically in $n$.
	
	These arguments are standard, and bounds of the same order are known for bounded
	kernels. For a bounded continuous kernel, Theorem~12 of
	\citet{rosasco2010learning} shows that, with high probability, the Nystr\"om
	extensions of the leading eigenvectors of the kernel matrix are close to the
	span of the leading eigenfunctions of the corresponding integral operator, with
	squared $L^2$ distance of order $n^{-1}$ divided by the squared eigengap.
	Theorem~\ref{thm:paper-mds} differs in two respects. First, it bounds the mean
	squared error of the embeddings themselves, rather than a distance between
	eigenspaces, with the same alignment $O_V$ for the objects in $\cD_2$ and for an
	independent object $V^\circ$. Second, it requires only $\E d_V(t_0,V)^2<\infty$
	and therefore allows unbounded metric spaces, for which the centered distance
	kernel $\kappa_V$ is unbounded.
	
	\subsection{Neural network approximation under low-dimensional structure}\label{sec:theory-covering}\label{sec:theory-lowdim}
	
	We next study the approximation and complexity of generative neural networks
	with low-dimensional structure. Combined with the MDS error bounds,
	these results will be used in Section~\ref{sec:theory-growing-rates}
	to verify the rate conditions of Theorem~\ref{cor:paper-null},
	even when the embedding dimensions increase with $n$.
	We assume that the oracle generator $g_{S,0}$ has components formed
	by composing smooth functions of a small number of variables,
	as in Equation~(13) of \citet{chen2026engressionerror}.
	The following definitions of smoothness and hierarchical composition are
	adapted from Definitions~2 and~3 in \citet{chai2026neural}.
	
	\begin{definition}[$(\beta,L_0)$-smooth functions]\label{def:paper-smoothness}
		Let $\beta=r+s$, where $r\in\bN_0$ and $0<s\le1$, and let $L_0>0$.
		For $F:\R^t\to\R$ with continuous partial derivatives through order $r$,
		define the norm
		\[
		\|F\|_{C^\beta(\R^t)}
		=\max\left\{
		\max_{|\nu|\le r}\sup_{x\in\R^t}|\partial^\nu F(x)|,\;
		\max_{|\nu|=r}\sup_{\substack{x,y\in\R^t\\x\ne y}}
		\frac{|\partial^\nu F(x)-\partial^\nu F(y)|}{\|x-y\|^s}
		\right\}.
		\]
		Here $\nu=(\nu_1,\ldots,\nu_t)\in\bN_0^t$,
		$|\nu|=\sum_{j=1}^t\nu_j$ and
		$\partial^\nu F=\partial^{|\nu|}F/(\partial x_1^{\nu_1}\cdots\partial x_t^{\nu_t})$.
		Write $C^\beta(\R^t)$ for the collection of functions with finite $\|\cdot\|_{C^\beta(\R^t)}$ norm.
		A function $F:\R^t\to\R$ is $(\beta,L_0)$-smooth if
		$\|F\|_{C^\beta(\R^t)}\le L_0$.
	\end{definition}
	
	\begin{definition}[Hierarchical composition]\label{def:paper-hierarchical-class}
		Fix $L_0>0$, $t^\ast\in\bN$ and a finite set
		$\cP\subset[1,\infty)\times[t^\ast]$ of pairs $(\beta,t)$.
		For $D\in\bN$, define the classes
		$\cH(D,k,\cP)$ on $\R^D$ recursively. Set
		$\cH(D,0,\cP)=\{x\mapsto x_j:j\in[D]\}$.
		For $k\ge1$, let
		\begin{equation}\label{eq:theory-hierarchical-class}
			\begin{aligned}
				\cH(D,k,\cP)=\bigl\{x\mapsto F\{h_1(x),\ldots,h_t(x)\}:\;&
				(\beta,t)\in\cP,h_j\in\cH(D,k-1,\cP),\ j\in[t]\bigr\},
			\end{aligned}
		\end{equation}
		where $F:\R^t\to\R$ is $(\beta,L_0)$-smooth.
	\end{definition}
	
	\begin{assumption}\label{ass:theory-lowdim-generator}\label{ass:theory-vector-smooth-generator}
		The generator $g_{S,0}$ in Section~\ref{sec:method-generators} can be
		chosen such that, for a sequence of positive integers $\ell_n$,
	$[g_{S,0}]_j\in\cH(p_Z+p_N,\ell_n,\cP)$ for $j\in[p_S]$,
	where $\beta>t/2+2$ for every $(\beta,t)\in\cP$.
	\end{assumption}
	
	By Assumption~\ref{ass:theory-lowdim-generator}, for each $j\in[p_S]$,
	$[g_{S,0}(w,u)]_j$ depends on at most
	$(t^\ast)^{\ell_n}$ entries of $(w^{\top},u^{\top})^{\top}$. Compositional assumptions of this type are standard in the analysis of
	deep ReLU networks for nonparametric regression
	\citep{schmidthieber2020nonparametric} and distributional regression
	\citep{chen2026engressionerror,chai2026neural}.
	To approximate $[g_{S,0}]_j$, we approximate each function $F$ in
	Equation~\eqref{eq:theory-hierarchical-class} and compose the resulting networks.

	To approximate the oracle generators, we consider a class of sparse
	feedforward ReLU networks. We also assume that the noise distribution
	has bounded support: $\supp(\bP_U)\subseteq[-R_0,R_0]^{p_N}$.
We adapt the network structure in Definition~1 of \citet{chai2026neural},
	which has a univariate output, to a $p_S$-dimensional output setting. For
	$S\in\{X,Y\}$, the network has $H_S\ge1$ hidden layers of common width
	$N_S\ge p_Z+p_N$. Write $d_{S,0}=p_Z+p_N$, $d_{S,k}=N_S$ for $k\in[H_S]$,
	and $d_{S,H_S+1}=p_S$ for the widths of the input, hidden and output
	layers. For $k\in[H_S+1]$, the $k$th layer has weight matrix
	$W_{S,k}\in\R^{d_{S,k}\times d_{S,k-1}}$ and bias vector
	$b_{S,k}\in\R^{d_{S,k}}$. We collect all the weights and biases of the
	network into a vector $\vartheta$ and assume all entries of $\vartheta$ are bounded by $\omega_S$ in absolute value and that
	$\|\vartheta\|_0\le d_{\mathrm{sp},S}$. We also bound the product of the operator norms of
	the weight matrices by $L_{W,S}$. Sparsity constraints are considered by
	\citet{chen2026engressionerror}, while operator-norm constraints are used
	to bound the Lipschitz constants of neural networks
	\citep{bartlett2017spectrally,miyato2018spectral}, which controls
	$L_{\mu,S}$. For an integer
	$1\le d_{\mathrm{sp},S}\le D_S$, $\omega_S>0$ and $L_{W,S}\in(0,\infty]$, let
	\begin{equation}\label{eq:theory-network-parameters}
		\begin{aligned}
			\Theta_S&=\left\{\vartheta\in[-\omega_S,\omega_S]^{D_S}:
			\|\vartheta\|_0\le d_{\mathrm{sp},S},\quad
			\prod_{k=1}^{H_S+1}\|W_{S,k}\|_{\mathrm{op}}\le L_{W,S}\right\},\\
			D_S&=N_S(p_Z+p_N+1)+(H_S-1)N_S(N_S+1)+p_S(N_S+1),
		\end{aligned}
	\end{equation}
	where $D_S$ is the total number of parameters.
	With the ReLU activation function $\sigma(t)=\max\{t,0\}$ acting entrywise,
	the network $g_\vartheta(w,u)$ with parameter $\vartheta$ is defined layer by layer as
	\begin{equation}\label{eq:theory-relu-network}
		\begin{aligned}
			h_{\vartheta,0}(w,u)&=(w^{\top},u^{\top})^{\top},\ \
			h_{\vartheta,k}(w,u)=\sigma\{W_{S,k}h_{\vartheta,k-1}(w,u)+b_{S,k}\},
			\ \  1\le k\le H_S,\\
			g_\vartheta(w,u)&=h_{\vartheta,H_S+1}(w,u)
			=W_{S,H_S+1}h_{\vartheta,H_S}(w,u)+b_{S,H_S+1}.
		\end{aligned}
	\end{equation}
	Take $\cG_{0,S}=\{g_\vartheta:\vartheta\in\Theta_S\}$ and form $\cG_S$
	as in Equation~\eqref{eq:theory-rotation-classes}.
	
	Write $\alpha_*=1/2+2/t^\ast$. Since $\beta>t/2+2$ and $t\le t^\ast$ for every
	$(\beta,t)\in\cP$, $\alpha_*$ is a lower bound for the ratios $\beta/t$, whose
	minimum over $\cP$ is the smoothness parameter $\gamma^\ast$ in Corollary~4.2 of
	\citet{chai2026neural}.
	The following proposition bounds $A_S$ and the constants $L_{\mu,S}$ and $v_S$ in
	Assumption~\ref{ass:theory-generator-class} for a suitable choice of the network
	parameters.
	
	\begin{proposition}\label{prop:paper-lowdim-approximation}
		Suppose Assumption~\ref{ass:theory-lowdim-generator} holds and
	$\supp(\bP_U)\subseteq[-R_0,R_0]^{p_N}$. For an integer $m\ge2$
		and a sufficiently large constant $C_0>1$, set the parameters of the
		network class $\cG_{0,S}$ to
		\begin{equation}\label{eq:theory-lowdim-network-choice}
			\begin{aligned}
				H_S&=2\ell_n-1, \ \
				L_{W,S}=\sqrt{p_S}C_0^{\ell_n}, \ \
				N_S=\max\{\lceil C_0^{\ell_n}p_Sm\rceil,p_Z+p_N\},\\
				\omega_S&=C_0m^{-1/2},\ \ \mbox{and}  \ \ d_{\mathrm{sp},S}=\min\{D_S,\lceil C_0^{\ell_n}p_S(p_Z+p_N+p_S)m\rceil\}.
			\end{aligned}
		\end{equation}
		Then we have
	\[
	A_S\le C C_0^{2\ell_n}p_S\log(em)m^{-(\alpha_*+1/2)}.
	\]
	Moreover, Assumption~\ref{ass:theory-generator-class}(\ref{eq:theory-mean-lipschitz})
	and~(\ref{eq:theory-generator-covering}) hold with $L_{\mu,S}$ and $v_S$ satisfying
	$L_{\mu,S}^2\le Cp_SC_0^{2\ell_n}$ and
	$v_S\le C\{C_0^{\ell_n}p_S(p_Z+p_N+p_S)m+p_S^2+p_Z^2\}$,
	and with $\Lambda_S$ given in Proposition~\ref{prop:paper-network-covering} in
		Appendix~\ref{app:proof-network-covering}.
		The constants $C_0$ and $C$ do not depend on $n$ or $m$.
	\end{proposition}
	
	For given $\ell_n$ and $p_S$, increasing $m$ improves the approximation
	ability of the network class without increasing its uniform Lipschitz constant,
	while $v_S$, and hence the stochastic error terms in
	Equation~\eqref{eq:theory-stochastic-rates}, increase with $m$.
	Thus, the approximation error
	$A_S$ can be reduced without amplifying the MDS embedding/Nystr\"om extension error
	$\rho_{Z,n}^2$ through the squared Lipschitz constant $L_{\mu,S}^2$.
	This is possible because each function $F$ in
	Equation~\eqref{eq:theory-hierarchical-class} is approximated by a shallow ReLU
	network with  $O(m)$ hidden nodes whose parameters are of order $m^{-1/2}$,
	which keeps the operator norms of its weight matrices bounded as $m$ grows.
	The rotations $O_S$ and $O_Z$ can make the input weights of $w$ and
	the output weights dense, which is accounted for by the factor
	$p_Z+p_N+p_S$ in $d_{\mathrm{sp},S}$.
	The terms $p_S^2$ and $p_Z^2$ in $v_S$ come from covering the orthogonal
	matrices $Q_S$ and $Q_Z$.
	In \citet{fan2024noise}, it is left as an open problem whether deep ReLU networks
	with a Lipschitz constant independent of their depth and width
	retain the approximation error of unconstrained networks.
	Proposition~\ref{prop:paper-lowdim-approximation} does not match their approximation
	rates for unconstrained networks, but it shows that a Lipschitz bound independent
	of the width still allows the approximation error to decrease polynomially in
	$m$.
	
	\subsection{Explicit rates with growing embedding dimensions}\label{sec:theory-growing-rates}
	
	The smoothness and hierarchical composition conditions in
	Assumption~\ref{ass:theory-lowdim-generator} alone do not ensure
	$\cE_S=o(n^{-1/2})$, since $\cE_S$ also contains MDS embedding/Nystr\"om
	extension errors and the conditioning truncation error $B_S$. We therefore impose further
	conditions on the eigenvalue and eigengaps in the oracle embedding, together with a bound on $B_S$.
	
	\begin{assumption}\label{ass:theory-spectral-truncation}
	\begin{compactenum}[(i)]
			\item\label{item:theory-spectral-decay}
			For each $V\in\{X,Y,Z\}$, the eigenvalues satisfy either polynomial decay,
			$\theta_{V,j}\asymp j^{-a_V}$ and $\gamma_{V,p_V}\gtrsim p_V^{-a_V-1}$
			with $a_V>1$, or exponential decay, $\theta_{V,j}\asymp e^{-b_Vj}$ and
			$\gamma_{V,p_V}\gtrsim e^{-b_Vp_V}$ with $b_V>0$.
			The comparison constants are independent of $n$ and $j$.
			\item\label{eq:theory-conditioning-tail-bound}
			For some $r>0$, $B_S\le C\sum_{j>p_Z}\theta_{Z,j}^{2r+1}$ for $S\in\{X,Y\}$.
	\end{compactenum}
\end{assumption}
	
	Assumption~\ref{ass:theory-spectral-truncation}(\ref{item:theory-spectral-decay})
	specifies how fast the spectral tails $\sum_{j>p_V}\theta_{V,j}$ vanish and
	how small the eigengap in
	Theorem~\ref{thm:paper-mds} can be. Its polynomial case is similar to the
	eigenvalue condition in Equation~(3.2) of \citet{hall2007methodology} for
	functional linear regression, which bounds every spacing
	$\theta_{V,j}-\theta_{V,j+1}$ from below by a multiple of $j^{-a_V-1}$;
	here only the eigengap at the truncation level $p_V$ is bounded.
	Assumption~\ref{ass:theory-spectral-truncation}(\ref{eq:theory-conditioning-tail-bound})
	controls how much the conditional mean embedding of $\phi_S(S)$ changes,
	on average, when we condition on $\phi_Z(Z)$ in place of $Z$.
	
	The following corollary gives rates for the network estimation error
	$A_S+\tau_S^2+J_S$ and the MDS embedding/Nystr\"om extension error
	$\rho_{S,n}^2+B_S+L_{\mu,S}^2\rho_{Z,n}^2$, together with conditions ensuring
	$\cE_S=o(n^{-1/2})$.

	\begin{corollary}\label{cor:paper-lowdim-rates}
		Suppose Assumptions~\ref{ass:theory-fourth-moment}--\ref{ass:theory-spectral-truncation} hold
	and $\supp(\bP_U)\subseteq[-R_0,R_0]^{p_N}$, with
		$\ell_n=o(\log n)$.
		Take $p_N\lesssim p_X+p_Y+p_Z$, $M_{\rm tr}\gtrsim n$ and $\delta_S=o(n^{-1/2})$.
		For the network class specified by Equation~\eqref{eq:theory-lowdim-network-choice},
		\begin{enumerate}[(i)]
			\item\label{item:theory-lowdim-polynomial}
			If the eigenvalues $\theta_{V,j}$ decay polynomially, take
			$p_V\asymp n^{\kappa_V}$
			and $m\asymp n^{(1-\kappa_{\max})/(\alpha_*+3/2)}$ where $\kappa_{\max}=\max_{V\in\{X,Y,Z\}}\kappa_V$. We have
			\begin{equation}\label{eq:theory-lowdim-polynomial-error-rates}
				\begin{aligned}
					A_S+\tau_S^2+J_S
					&\le Cn^{-\frac{(\alpha_*+1/2)(1-\kappa_{\max})-(\alpha_*+3/2)\kappa_S}{\alpha_*+3/2}+o(1)}
					+Cn^{-1+2\kappa_Z+o(1)},\\
					\rho_{S,n}^2+B_S+L_{\mu,S}^2\rho_{Z,n}^2
					&\le Cn^{-\min\{1-(2a_S+2)\kappa_S,\;\kappa_Z[a_Z(2r+1)-1],\;1-\kappa_S-(2a_Z+2)\kappa_Z\}+o(1)}
				\end{aligned}
			\end{equation}
			and $\cE_S=o(n^{-1/2})$ holds if $\kappa_S$ and $\kappa_Z$ are suitably chosen
			(see Equation~\eqref{eq:theory-lowdim-polynomial-choice} in
			Appendix~\ref{app:proof-deep-approximation}).
			
			\item\label{item:theory-lowdim-exponential}
			If the eigenvalues $\theta_{V,j}$ decay exponentially, take
			$\underline\kappa_V\log n\le p_V\le\overline\kappa_V\log n$
			for all sufficiently large $n$ and
			$m\asymp n^{1/(\alpha_*+3/2)}$,
			where $0<\underline\kappa_V\le\overline\kappa_V$ are constants.
			We have
			\begin{equation}\label{eq:theory-lowdim-exponential-error-rates}
				\begin{aligned}
					A_S+\tau_S^2+J_S
					&\le Cn^{-\frac{\alpha_*+1/2}{\alpha_*+3/2}+o(1)},\\
					\rho_{S,n}^2+B_S+L_{\mu,S}^2\rho_{Z,n}^2
					&\le Cn^{-\min\{1-2b_S\overline\kappa_S,\;b_Z(2r+1)\underline\kappa_Z,\;1-2b_Z\overline\kappa_Z\}+o(1)}
				\end{aligned}
			\end{equation}
			and $\cE_S=o(n^{-1/2})$ holds if $\underline\kappa_V$ and $\overline\kappa_V$
			are suitably chosen (see Equation~\eqref{eq:theory-lowdim-exponential-choice} in
			Appendix~\ref{app:proof-deep-approximation}).
		\end{enumerate}
		For fixed $\kappa_V$ satisfying Equation~\eqref{eq:theory-lowdim-polynomial-choice} or fixed
		$\underline\kappa_V,\overline\kappa_V$ satisfying Equation~\eqref{eq:theory-lowdim-exponential-choice},
		the conclusion $\cE_S=o(n^{-1/2})$ also holds if $\ell_n\le c\log n$ for a sufficiently small $c>0$.
	\end{corollary}
	
	To compare with the approximation and stochastic error bounds in \cite{chai2026neural}, suppose the embedding dimension $p_V$ is fixed (i.e., $\kappa_V=0$). Then Corollary~\ref{cor:paper-lowdim-rates} implies
	\begin{equation}\label{eq:theory-fixed-dimension-network-rate}
		A_S+\tau_S^2+J_S
		\le Cn^{-(\alpha_*+1/2)/(\alpha_*+3/2)+o(1)}.
	\end{equation}
	The exponent $(\alpha_*+1/2)/(\alpha_*+3/2)$ is smaller than
	$2\gamma^\ast/(2\gamma^\ast+1)$ in Corollary~4.2 of \citet{chai2026neural}.
	This is because, apart from the approximation error $A_S$ and the stochastic
	error $\tau_S^2+J_S$, the combined error $\cE_S$ also includes MDS
	embedding/Nystr\"om extension errors. Specifically, the term
	$L_{\mu,S}^2\rho_{Z,n}^2$ depends on both the uniform Lipschitz constant
	$L_{\mu,S}$ and $\rho_{Z,n}^2$.
	Enlarging the network class can reduce $A_S$, but may also increase $L_{\mu,S}$.
	To balance these two terms, Proposition~\ref{prop:paper-lowdim-approximation} uses
	$\omega_S=C_0m^{-1/2}$ and the operator-norm product bound
	$L_{W,S}=\sqrt{p_S}C_0^{\ell_n}$, which gives
	$L_{\mu,S}^2\le Cp_SC_0^{2\ell_n}$ independently of $m$.
	Thus increasing $m$ improves network approximation without increasing $L_{W,S}$.
	
	When $Z\in\R^{p_Z}$ is a random vector, the term
	$L_{\mu,S}^2\rho_{Z,n}^2$ disappears and we can remove the sparsity and operator-norm restrictions on our network class by taking
	$d_{\mathrm{sp},S}=D_S$ and $L_{W,S}=\infty$ and allowing the parameter bound $\omega_S$ to grow
	with the network width. In this case and with fixed $p_V$, we obtain
	$A_S+\tau_S^2+J_S\le Cn^{-2\alpha_*/(2\alpha_*+1)+o(1)}$, the same rate as in
	Corollary~4.2 of \citet{chai2026neural} with $\gamma^\ast$ replaced by its
	lower bound $\alpha_*$; see Appendix~\ref{app:vector-network-rates} for more details.

	\section{Simulation Studies}\label{sec:simulation}
	
	We evaluate the empirical size and power of the proposed test, comparing it with the distance profile embedding (DPE) test of \citet{tan2026dpe} and the spectral generalized covariance measure (SGCM) test of \citet{miyazaki2025sgcm}. When $Z$ is a random object, both methods estimate nuisance quantities by kernel ridge regression with regularization parameter $\varepsilon$. When $Z$ is a vector, DPE still uses kernel ridge regression, whereas SGCM allows other regression methods, including XGBoost and neural networks. We report empirical size at the nominal 5\% level and, following \citet{dominguez2000size}, size-adjusted power under alternatives, using the 5\% empirical quantile of null p-values as the rejection threshold for each method.
	
	For our test, we use the Gaussian kernel $f_V(r)=\exp(-r/h_V)$ in Equation~\eqref{eq:method-full-kernels}. For $\wh T_1$, we set $h_S$ to twice the median of $\{\|\wt S_i-\wt S_j\|^2:i,j\in\cJ_1,i\ne j\}$ and $h_Z$ to twice the median of $\{d_Z(Z_i,Z_j):i,j\in\cJ_1,i\ne j\}$, following the median heuristic \citep[Section~8]{gretton2012kernel}. The neural generators are trained by minimizing the empirical risk in Equation~\eqref{eq:method-sampled-loss}; network architectures and training details are given in Appendix~\ref{app:simulation-settings}. We take $M_{\rm ev}=10$ and $B=1000$.
	
	\subsection{Distribution-valued conditioning variable}\label{sec:simulation-distributional}
	
	We first consider Gaussian-distribution-valued $X$, $Y$, and $Z$ equipped with the 2-Wasserstein distance. For an integer $d\ge1$ and $i\in[2n]$, let $\zeta_i\sim N(0,I_d)$, $\eta_{1i},\eta_{2i}\sim{\rm Unif}[-1,1]$, $\eta_{3i}\sim N(0,1)$, and $\Delta_i\sim{\rm Bernoulli}(p)$ be mutually independent. Setting $a_i=d^{-1/2}\ones_d^{\top}\zeta_i$ and $c_i=\Delta_i\eta_{1i}+(1-\Delta_i)\eta_{2i}$, we generate
	\[
	Z_i=N(\zeta_i,I_d),\qquad X_i=N(a_i+\eta_{1i},1),\qquad Y_i=N\{a_i+\sin(4\pi c_i)+0.6\eta_{3i},1\}.
	\]
	Here $p=0$ corresponds to $H_0$. Since $c_i\sim{\rm Unif}[-1,1]$ for every $p$, the conditional marginal distributions of $X$ and $Y$ given $Z$ do not depend on $p$. We set $n=1000$, $d\in\{1,5\}$, and $p\in\{0,0.1,0.2,0.3,0.5,1\}$, with 1000 replications under $H_0$ and 500 under each alternative.
	
	For our test, we consider $p_X=p_Y=p_Z\in\{20,50\}$ and a data-driven choice of each $p_V$: the smallest integer for which the leading $p_V$ eigenvalues of $B_V$ in Equation~\eqref{eq:method-mds-matrices} account for at least 85\% of its trace. This rule selects $p_X,p_Y\in\{5,6\}$, $p_Z\in\{5,6\}$ when $d=1$, and $p_Z$ between 197 and 205 when $d=5$. For DPE, we use either the empirical distribution of the $2n$ observed objects as the reference measure or a Monte Carlo reference measure of random Gaussian distributions adapted from Scenario~3 of \citet[Section~7.3]{tan2026dpe}. For both DPE and SGCM, we consider $\varepsilon\in\{10^{-5},10^{-4},5\times10^{-4},1.5\times10^{-3},5\times10^{-3}\}$; the last value is used in \citet{tan2026dpe}. The other settings of the two methods are given in Appendix~\ref{app:simulation-settings}.
	
	\begin{table}[!ht]
		\centering
		\begin{tabular}{@{}cllrrrrrr@{}}
			\toprule
			& & & Size & \multicolumn{5}{c}{Size-adjusted power}\\
			\cmidrule(lr){4-4}\cmidrule(l){5-9}
			$d$ & Method & Setting & $p=0$ & $p=0.1$ & $p=0.2$ & $p=0.3$ & $p=0.5$ & $p=1$\\
			\midrule
			\multirow{18}{*}{1} & DPE (empirical) & $\varepsilon=10^{-5}$ & 4.8 & 9.2 & 13.8 & 23.4 & 65.0 & 100.0\\
			& & $\varepsilon=10^{-4}$ & 4.4 & 9.4 & 14.2 & 24.8 & 66.2 & 100.0\\
			& & $\varepsilon=5\times10^{-4}$ & 4.2 & 9.6 & 14.8 & 25.0 & 66.0 & 100.0\\
			& & $\varepsilon=1.5\times10^{-3}$ & 4.2 & 9.6 & 14.8 & 25.4 & 66.2 & 100.0\\
			& & $\varepsilon=5\times10^{-3}$ & 4.0 & 9.4 & 14.6 & 25.4 & 65.6 & 100.0\\
			\cmidrule(l){2-9}
			& DPE (Gaussian) & $\varepsilon=10^{-5}$ & 4.1 & 9.2 & 13.6 & 22.6 & 62.8 & 100.0\\
			& & $\varepsilon=10^{-4}$ & 4.0 & 9.2 & 13.6 & 22.8 & 63.4 & 100.0\\
			& & $\varepsilon=5\times10^{-4}$ & 3.8 & 9.2 & 14.0 & 23.0 & 63.2 & 100.0\\
			& & $\varepsilon=1.5\times10^{-3}$ & 3.7 & 9.2 & 14.4 & 22.8 & 62.8 & 100.0\\
			& & $\varepsilon=5\times10^{-3}$ & 3.6 & 9.2 & 14.0 & 21.6 & 62.4 & 100.0\\
			\cmidrule(l){2-9}
			& SGCM & $\varepsilon=10^{-5}$ & 37.8 & 6.2 & 7.6 & 11.0 & 30.2 & 93.4\\
			& & $\varepsilon=10^{-4}$ & 10.4 & 6.6 & 9.0 & 14.0 & 40.2 & 97.6\\
			& & $\varepsilon=5\times10^{-4}$ & 6.8 & 8.0 & 10.2 & 14.0 & 43.0 & 98.0\\
			& & $\varepsilon=1.5\times10^{-3}$ & 5.2 & 7.4 & 10.4 & 14.2 & 44.2 & 98.2\\
			& & $\varepsilon=5\times10^{-3}$ & 5.0 & 7.2 & 10.0 & 14.2 & 43.8 & 98.4\\
			\cmidrule(l){2-9}
			& $\wh T_n$ & $p_V=20$ & 5.9 & 6.8 & 11.0 & 15.8 & 38.6 & 98.4\\
			& & $p_V=50$ & 7.0 & 7.6 & 12.4 & 16.6 & 42.0 & 98.8\\
			& & 85\% rule & 5.5 & 7.4 & 9.2 & 14.8 & 41.2 & 98.2\\
			\midrule
			\multirow{18}{*}{5} & DPE (empirical) & $\varepsilon=10^{-5}$ & 15.9 & 9.0 & 9.4 & 20.2 & 48.8 & 99.8\\
			& & $\varepsilon=10^{-4}$ & 10.4 & 7.8 & 8.4 & 17.2 & 42.6 & 99.2\\
			& & $\varepsilon=5\times10^{-4}$ & 11.5 & 5.0 & 5.2 & 7.8 & 27.2 & 97.6\\
			& & $\varepsilon=1.5\times10^{-3}$ & 29.0 & 4.0 & 1.4 & 1.2 & 4.2 & 80.4\\
			& & $\varepsilon=5\times10^{-3}$ & 93.3 & 2.2 & 0.8 & 0.2 & 0.0 & 4.4\\
			\cmidrule(l){2-9}
			& DPE (Gaussian) & $\varepsilon=10^{-5}$ & 13.4 & 9.8 & 10.2 & 18.6 & 44.4 & 98.8\\
			& & $\varepsilon=10^{-4}$ & 9.3 & 7.8 & 8.2 & 16.2 & 41.0 & 98.8\\
			& & $\varepsilon=5\times10^{-4}$ & 10.2 & 5.4 & 6.0 & 7.2 & 21.6 & 96.2\\
			& & $\varepsilon=1.5\times10^{-3}$ & 23.9 & 4.0 & 2.6 & 1.2 & 3.4 & 72.0\\
			& & $\varepsilon=5\times10^{-3}$ & 83.8 & 3.6 & 1.4 & 0.2 & 0.0 & 2.0\\
			\cmidrule(l){2-9}
			& SGCM & $\varepsilon=10^{-5}$ & 65.9 & 4.8 & 6.6 & 8.0 & 22.0 & 84.6\\
			& & $\varepsilon=10^{-4}$ & 40.6 & 5.6 & 6.6 & 8.8 & 26.4 & 91.6\\
			& & $\varepsilon=5\times10^{-4}$ & 16.1 & 4.4 & 4.8 & 6.8 & 21.4 & 91.8\\
			& & $\varepsilon=1.5\times10^{-3}$ & 14.4 & 1.8 & 1.2 & 2.4 & 6.2 & 72.4\\
			& & $\varepsilon=5\times10^{-3}$ & 58.6 & 0.0 & 0.0 & 0.0 & 0.0 & 0.0\\
			\cmidrule(l){2-9}
			& $\wh T_n$ & $p_V=20$ & 6.4 & 5.0 & 10.0 & 14.8 & 42.8 & 98.8\\
			& & $p_V=50$ & 6.6 & 6.0 & 9.6 & 13.4 & 43.0 & 99.8\\
			& & 85\% rule & 6.3 & 6.2 & 10.2 & 14.0 & 42.0 & 99.0\\
			\bottomrule
		\end{tabular}
		\caption{Empirical size and size-adjusted power (in \%) at nominal level 5\% for the distributional data in Section~\ref{sec:simulation-distributional}.}
		\label{tab:simulation-distributional}

	\end{table}
	
	Table~\ref{tab:simulation-distributional} shows that our test has empirical sizes of 5.5--7.0\% for $d=1$ and 6.3--6.6\% for $d=5$, with relatively little variation across choices of $p_V$. DPE is slightly conservative for $d=1$ (3.6--4.8\%) but is oversized for $d=5$, with sizes ranging from 9.3\% to 93.3\% depending on $\varepsilon$. SGCM is also sensitive to $\varepsilon$: its size ranges from 5.0\% to 37.8\% for $d=1$ and from 14.4\% to 65.9\% for $d=5$.
	
	The three choices of $p_V$ also yield similar size-adjusted power, exceeding 98\% at $p=1$ in both settings. For $d=1$, DPE has higher power than our test (62.4--66.2\% against 38.6--42.0\% at $p=0.5$), and SGCM with $\varepsilon\ge5\times10^{-4}$ has comparable power. For $d=5$, DPE with $\varepsilon=10^{-4}$ has comparable power, but its power falls below 5\% at all alternatives with $p\le0.5$ when $\varepsilon\ge1.5\times10^{-3}$, and SGCM has lower power than our test for every $\varepsilon$ when $p\ge0.2$ (at most 26.4\% against 42.0--43.0\% at $p=0.5$).
	
\begin{figure}[!ht]
		\centering
		\includegraphics[width=0.82\textwidth]{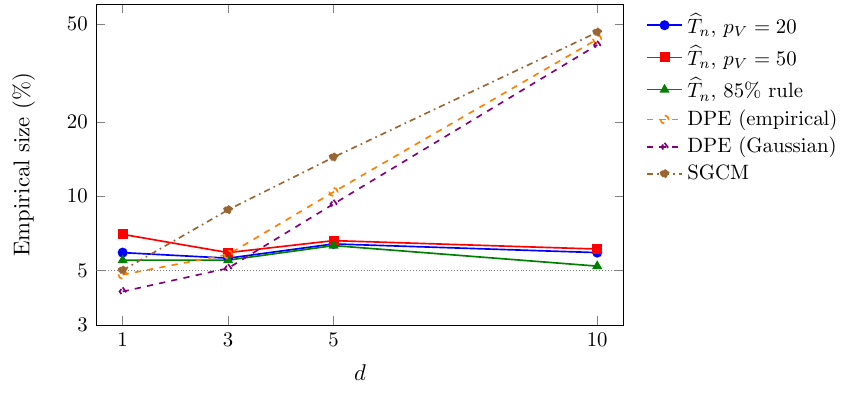}
		\caption{Empirical size (in \%, logarithmic scale) against $d$ at nominal level 5\% for the distributional data in Section~\ref{sec:simulation-distributional}. For DPE and SGCM, each point uses the value of $\varepsilon$ whose size is closest to 5\% at that $d$. The dotted line marks the nominal level.}
		\label{fig:simulation-size-dimension}
	\end{figure}

	To further assess the effect of the dimension of $Z$ on size, we include $d=3$ and $d=10$ under $H_0$. Figure~\ref{fig:simulation-size-dimension} compares the methods, selecting for DPE and SGCM the value of $\varepsilon$ that gives empirical size closest to 5\% at each $d$. Across these dimensions, the size of our test remains between 5.2\% and 7.0\%, while that of the competing methods increases with $d$. At $d=10$, the sizes are 43.4\% and 41.1\% for DPE with the empirical and Gaussian reference measures, respectively, and 46.3\% for SGCM.
	
	\subsection{High-dimensional Euclidean conditioning variable}\label{sec:simulation-vector-z}
	
	We next consider distribution-valued $X$ and $Y$ with a 100-dimensional Euclidean conditioning variable $Z$. For $i\in[2n]$, let $Z_i\sim N(0,I_{100})$, $\eta_{1i},\eta_{2i}\sim N(0,1)$, and $\Delta_i\sim{\rm Bernoulli}(p)$ be mutually independent. With $a_i=100^{-1/2}\ones_{100}^{\top}Z_i$, we generate
	\[
	X_i=N\{a_i+(1-\Delta_i)\eta_{1i}+\Delta_i\eta_{2i},1\},\qquad Y_i=N(a_i+\eta_{2i},1).
	\]
	We equip $X$ and $Y$ with the 2-Wasserstein distance and $Z$ with the Euclidean distance. As before, $p=0$ corresponds to $H_0$, and the conditional marginal distributions of $X$ and $Y$ given $Z$ are unchanged as $p$ varies.
	
	We use $Z$ directly, as in Section~\ref{sec:method-vector-z}, setting $\wt Z_i=Z_i$ and $p_Z=100$. We take $n=1000$ and $p\in\{0,0.05,0.1,0.15,0.2,0.4\}$, with 1000 replications under $H_0$ and 500 under each alternative. For our test, we use $p_X=p_Y\in\{20,50\}$ or the 85\% rule from Section~\ref{sec:simulation-distributional}, which selects $p_X,p_Y\in\{5,6\}$. For DPE, we use the same grid of $\varepsilon$ and two types of reference measures: the empirical distribution of the observed objects, or a Monte Carlo reference measure. For SGCM, the conditional expectations of eigenfunctions given $Z$ are estimated by XGBoost \citep{chen2016xgboost}. The details are given in Appendix~\ref{app:simulation-settings}.
	
	\begin{table}[!ht]
		\centering
		\begin{tabular}{@{}llrrrrrr@{}}
			\toprule
			& & Size & \multicolumn{5}{c}{Size-adjusted power}\\
			\cmidrule(lr){3-3}\cmidrule(l){4-8}
			Method & Setting & $p=0$ & $p=0.05$ & $p=0.1$ & $p=0.15$ & $p=0.2$ & $p=0.4$\\
			\midrule
			DPE (empirical) & $\varepsilon=10^{-5}$ & 100.0 & 67.8 & 99.2 & 100.0 & 100.0 & 100.0\\
			& $\varepsilon=10^{-4}$ & 100.0 & 70.4 & 99.6 & 100.0 & 100.0 & 100.0\\
			& $\varepsilon=5\times10^{-4}$ & 100.0 & 75.6 & 99.6 & 100.0 & 100.0 & 100.0\\
			& $\varepsilon=1.5\times10^{-3}$ & 100.0 & 79.4 & 100.0 & 100.0 & 100.0 & 100.0\\
			& $\varepsilon=5\times10^{-3}$ & 100.0 & 76.8 & 100.0 & 100.0 & 100.0 & 100.0\\
			\midrule
			DPE (Gaussian) & $\varepsilon=10^{-5}$ & 100.0 & 59.2 & 98.6 & 100.0 & 100.0 & 100.0\\
			& $\varepsilon=10^{-4}$ & 100.0 & 65.6 & 99.8 & 100.0 & 100.0 & 100.0\\
			& $\varepsilon=5\times10^{-4}$ & 100.0 & 72.4 & 99.8 & 100.0 & 100.0 & 100.0\\
			& $\varepsilon=1.5\times10^{-3}$ & 100.0 & 75.6 & 100.0 & 100.0 & 100.0 & 100.0\\
			& $\varepsilon=5\times10^{-3}$ & 100.0 & 70.6 & 99.8 & 100.0 & 100.0 & 100.0\\
			\midrule
			SGCM & XGBoost & 100.0 & -- & -- & -- & -- & --\\
			\midrule
			$\wh T_n$ & $p_V=20$ & 7.6 & 23.2 & 73.2 & 97.2 & 100.0 & 100.0\\
			& $p_V=50$ & 7.8 & 18.4 & 63.2 & 95.8 & 99.8 & 100.0\\
			& 85\% rule & 7.5 & 16.6 & 61.0 & 95.0 & 99.6 & 100.0\\
			\bottomrule
		\end{tabular}
		\caption{Empirical size and size-adjusted power (in \%) at nominal level 5\% for the Euclidean conditioning variable in Section~\ref{sec:simulation-vector-z}.}
		\label{tab:simulation-vector-z}

	\end{table}
	
	As shown in Table~\ref{tab:simulation-vector-z}, our test has empirical sizes of 7.5--7.8\%, whereas both DPE (under either reference measure and all values of $\varepsilon$) and SGCM reject in every null replication. When the sample size is increased to $n=2000$, with 1000 null replications, the sizes of our test decrease to 6.3\%, 6.1\%, and 5.8\% for $p_V=20$, $p_V=50$, and the 85\% rule, respectively, while SGCM continues to reject in every null replication.
	
	Among the three embedding choices, $p_V=20$ gives the highest size-adjusted power for our test; all three attain at least 95\% at $p=0.15$ and at least 99.6\% at $p\ge0.2$. DPE has higher size-adjusted power at $p=0.05$ and $p=0.1$, but its severe size distortion makes these comparisons difficult to interpret. Because SGCM produces p-values of zero in every replication, its size-adjusted power cannot be computed.
	
	Overall, our test shows more stable empirical size across the settings considered, with relatively little sensitivity to the embedding dimension. In contrast, DPE and SGCM exhibit substantial size distortions when $Z$ is a high-dimensional vector or a higher-dimensional distribution, and the results for object-valued $Z$ are sensitive to the ridge regularization parameter.
	
	\section{Conclusion}\label{sec:conclusion}

	We have developed a test of conditional independence between random objects $X$ and $Y$ given a random object or a high-dimensional Euclidean vector $Z$. Our approach uses MDS and Nystr\"om extensions to obtain finite-dimensional representations of object-valued data, and then learns the conditional distributions in the embedded space using neural generators. These estimates are incorporated into a doubly robust MMDCI statistic.
	
	A main challenge in this approach is understanding how the estimation and truncation of Hilbert embeddings affect the validity of the test. In particular, errors in estimating the embedding of $Z$ are amplified by the Lipschitz constant of the generated conditional mean embeddings, creating a trade-off between the approximation accuracy of the generators and their sensitivity to embedding errors. To address this, we establish error bounds for MDS and Nystr\"om extensions under finite fourth moments and construct sparse ReLU networks whose approximation error decreases with width while their Lipschitz constant remains controlled. These results provide conditions for asymptotically valid inference even when the embedding dimensions grow with the sample size. Our simulations also show relatively stable empirical size across the settings considered.
	
	Our analysis leaves several questions for future work. One is how to select the embedding dimensions in a data-driven manner while retaining theoretical guarantees. More broadly, the proposed framework may be extended to other hypothesis testing problems for random objects, such as testing equality of distributions or conditional distributions, as well as to dependent object-valued data and metric spaces beyond negative type.

\paragraph{Supplementary material.} The supplement contains the proofs of the theoretical results, the covering number bounds for the network classes, the rates for Euclidean conditioning variables, and the implementation details of the simulations.

\nocite{Dubey2024,Lundborg2022,ShahPeters2020,TangLi2026,Zhang2011,ZhouMuller2026,ariascastro2020perturbation,bartlett2017spectrally,bellot2019gcit,boucheron2013concentration,candes2018panning,chai2026neural,chen2016xgboost,chen2020distribution,chen2026engressionerror,daviskahan1970rotation,devito2013mercer,dinezza2012hitchhiker,dominguez2000size,e2020banach,e2022barron,fan2024factor,fan2024noise,feng2017sparse,folland1999real,fukumizu2007kernel,golubvanloan2013,gower1968adding,gretton2012kernel,gyorfi2002distribution,hall2007methodology,horvath2012inference,hsing2015theoretical,huang2022kpc,huang2023highdimensional,iao2025deep,jiao2023approximation,kallenberg2002foundations,kechris1995classical,klusowski2018approximation,lim2024classical,lyons2013,lyons2021second,markus1964eigen,miyato2018spectral,miyazaki2025sgcm,moertersperes2010brownian,mourier1953elements,pearl2014probabilistic,petersen2019frechet,rauhut2010compressive,rosasco2010learning,schmidthieber2020nonparametric,schmidtvu2024correction,serfling1980approximation,shen2025engression,sheng2023distance,shi2021dgcit,tan2026dpe,torgerson1952multidimensional,yuan2007dimension,zhang2024doubly,zhangzhushao2025,zhou2022deep,zhou2026e2m,ziegel2024characteristic}
{\fontsize{11.15}{11.15}\selectfont
\setlength{\bibsep}{0pt plus 0.24ex}
\bibliographystyle{chicago}
\bibliography{Bib_ObjCI_arXiv}
}
\end{document}